\documentclass[twocolumn, journal]{IEEEtran}
\IEEEoverridecommandlockouts

\usepackage{color}
\usepackage{bm}
\usepackage{graphicx}
\usepackage{amsmath}
\usepackage{amssymb}
\usepackage{algorithm}
\usepackage{algorithmic}
\usepackage{amsmath}
\usepackage{multirow}
\usepackage{booktabs}
\usepackage{array}
\usepackage{amsthm}
\usepackage{lipsum}
\usepackage{enumerate}
\usepackage{stfloats}
\usepackage{subfigure}
\usepackage{cases}

\newcommand{\be}{\begin{equation}}
\newcommand{\ee}{\end{equation}}
\newcommand{\bea}{\begin{eqnarray}}
\newcommand{\eea}{\end{eqnarray}}
\newcommand{\ba}{\begin{array}}
\newcommand{\ea}{\end{array}}

\newcommand{\non}{\nonumber}

\title{Intelligent Reflecting Surface Enhanced Wideband MIMO-OFDM Communications: From Practical Model to Reflection Optimization}

\author{Hongyu Li,~\IEEEmembership{Student Member,~IEEE,}
       Wenhao Cai,
       Yang Liu,~\IEEEmembership{Member,~IEEE,}
       Ming Li,~\IEEEmembership{Senior Member,~IEEE,}\\
       Qian Liu,~\IEEEmembership{Member,~IEEE,}
       and Qingqing Wu,~\IEEEmembership{Member,~IEEE}

\thanks{H. Li, W. Cai, Y. Liu, and M. Li are with the School of Information and Communication Engineering, Dalian University of Technology, Dalian 116024, China, (e-mail: hongyuli@mail.dlut.edu.cn, wenhaocai@mail.dlut.edu.cn, yangliu613@dlut.edu.cn, mli@dlut.edu.cn).}
\thanks{Q. Liu is with the School of Computer Science and Technology, Dalian University of Technology, Dalian 116024, China (e-mail: qianliu@dlut.edu.cn).}
\thanks{Q. Wu is with the State Key Laboratory of Internet of Things for Smart City, University of Macau, Macau 999078, China (e-mail: qingqingwu@um.edu.mo).}
}

\begin{document}

\maketitle
\thispagestyle{empty}
\begin{abstract}
Intelligent reflecting surface (IRS) is envisioned as a revolutionary technology for future wireless communication systems since it can intelligently change radio environment and integrate it into wireless communication optimization.
However, most existing works adopted an ideal IRS reflection model, which is impractical and can cause significant performance degradation in realistic wideband systems.
To address this issue, we first study the \textit{dual phase- and amplitude-squint} effect of reflected signals and present a simplified practical IRS reflection model for wideband signals.
Then, an IRS enhanced wideband multiuser multi-input single-output orthogonal frequency division multiplexing (MU-MISO-OFDM) system is investigated.
We aim to jointly design the transmit beamformer and IRS reflection for the case of using both continuous and discrete phase shifters to maximize the average sum-rate over all subcarriers.
By exploiting the relationship between sum-rate maximization and mean square error (MSE) minimization, the original problem is equivalently transformed into a multi-block/variable problem, which can be efficiently solved by the block coordinate descent (BCD) method. Complexity and convergence for both cases are analyzed or illustrated. Simulation results demonstrate that the proposed algorithm can offer significant average sum-rate enhancement compared to that achieved using the ideal IRS reflection model, which confirms the importance of the use of the practical model for the design of wideband systems.
\end{abstract}


\section{Introduction}

The continuous popularizing of intelligent devices and the rapid development of emerging wireless services have spurred the exponential increase of the demand for wireless network traffic.
This motivates the research on key enabling technologies, such as massive multi-input multi-output (MIMO), ultra-dense network, and the use of millimeter wave (mmWave) bands \cite{Lee CM 2014}-\cite{Q Wu 2017}, for the fifth-generation (5G) and beyond networks.
However, the above technologies still inevitably face challenges mainly due to high cost and power consumptions when employing multiple antennas, cells (base stations (BSs)), and/or hardware components (e.g. radio frequency (RF) chains).
Therefore, researchers have never stopped their efforts to seek promising solutions to accommodate the demanding data rate and diverse quality of service (QoS) requirements for future wireless communications.

In the current paradigm of wireless communication, the radio environment and wireless propagation medium remain an uncontrollable factor, which cannot be included in the optimization formulations.
Thus, channel fading effect due to the randomness in the radio environment is generally a major challenge for the maximization of energy- and/or spectral-efficiency (EE/SE) performance of wireless communications. Recently, an innovative concept of intelligent reflecting surface (IRS) has been introduced in the wireless communication research community as a revolutionary technology, which can combat stochastic wireless propagation medium and achieve controllable radio environment \cite{Q Wu 2019}-\cite{Y Liu 2020}.

The IRS consists of a large number of nearly passive elements with ultra-low power consumption.
Particularly, each element of IRS is composed of configurable electromagnetic (EM) internals, which are capable of controlling the phase shift and amplitude of the incident EM wave in a programmable manner.
Adaptively adjusting elements of IRS can collaboratively achieve
reflection beamforming and shape the propagation environment suitable for wireless communications.
In this way, the channel/beamforming gain can be effectively improved and the communication quality can be enhanced.
Free of containing radio frequency (RF) chains, large-scale IRS can be deployed in different communication scenarios with substantially reduced power consumption and cost.
Therefore, IRS is envisioned to revolutionize the current communication optimization paradigm by integrating the smart radio environment and recently considered as a key technology in six-generation (6G) wireless networks \cite{N Rajatheva}.

Attracted by its sheer advantage, the investigation of IRS for improving the performance of various wireless communication systems is a thriving research area recently.
A majority of recent research efforts have been devoted to the IRS designs with focus on power allocation and/or beamformer for both single-user systems \cite{X Yu 2019}-\cite{Y Han} and multi-user systems \cite{C Huang}-\cite{M Li 2020} using different metrics, e.g. power minimization \cite{R Zhang}, \cite{J Zhao}, max-min fairness \cite{J Zhao}, \cite{M Li 2020}, SE maximization \cite{X Yu 2019}-\cite{Y Han}, \cite{H Guo}, and EE maximization \cite{C Huang}.
In some recent works \cite{B Di}-\cite{J Xu}, practical IRS implementation with finite/low-resolution phase shifts are considered.
To fully reap the performance improvement promised by IRS, many researchers also studied the coordination of multiple IRSs \cite{J He}-\cite{B Zheng Multi-IRS}.
Moreover, IRS technique has also been applied in various other applications, e.g., physical layer security \cite{M Cui}-\cite{D Xu}, cognitive radio \cite{L Zhang}-\cite{X Guan}, index modulation \cite{A Khaleel}, \cite{E Basar}, as well as multiple access system \cite{B Zheng NOMA}, etc.

It is worth noting that the IRS-assisted wireless communication systems mentioned above are restricted to narrowband channels.
When considering more general wideband frequency-selective channels,
the problem will be quite different and more difficult to be solved since the common IRS reflection pattern should be designed for all subcarriers, while the conventional digital beamformers can be individually optimized for each subcarrier.
Limited work has studied the IRS-enhanced wideband orthogonal frequency division multiplexing (OFDM) system for both simple single-input single-output (SISO) case \cite{Y Yang}-\cite{T Bai} and more typical multi-user MISO case \cite{H Li}.

The aforementioned work assumes an \textit{ideal} IRS reflection model, i.e., each element has constant magnitude, variable phase shift, and the same response for wideband signals.
The design of IRS under such an ideal reflection model can be easily implemented using classical optimization tools, e.g., semidefinite relaxation (SDR), manifold optimization, and majorization minimization (MM), etc.
However, it is practically difficult to implement an IRS having such an ideal reflection model due to the hardware circuit limitation \cite{H Rajagopalan}, \cite{W Tang}.
Therefore, these ``ideal'' designs will cause non-negligible performance loss in realistic systems since the ideal model cannot precisely capture the reflection response of a practical IRS.
It is thus important and necessary to analyze the response characteristic of a practical IRS and establish an accurate and practical IRS reflection model.
The authors in \cite{S Abeywickrama} have illustrated the fundamental relationship between reflection amplitude and phase shift under a narrowband scenario and demonstrated the performance enhancement with their proposed practical model compared to that with the ideal one.
When extending to wideband communications, unfortunately, the above two-dimensional amplitude-phase relationship cannot accurately describe the response of the practical IRS since the amplitude and phase shift will vary with the frequencies of incident signals, an effect referred to as \textit{dual phase- and amplitude-squint}.
In our previous work \cite{W Cai}, we have analyzed this issue and established a three-dimensional amplitude-frequency-phase relationship to precisely describe the dual phase- and amplitude-squint effect of practical IRS in wideband systems. Nevertheless, this practical model is so complicated that it will cause great difficulties in the IRS reflection design. This motivates us to further simplify the practical IRS reflection model in order to facilitate the reflection design without significant accuracy loss.

In this paper, we consider an IRS-enhanced wideband MU-MISO-OFDM communication system. Specifically, we present a simplified practical reflection model of IRS and take it into consideration for the reflection design. Our main contributions are summarized as follows:
\begin{itemize}
\item We analyze the characteristic of IRS elements, i.e., reflection phase and amplitude variations of IRS elements when responding to signals with different frequencies, a phenomenon referred to as dual phase- and amplitude-squint. Based on our previous work, we present a leaner practical model of IRS reflection, which can be widely applied in the designs of typical communication scenarios.
\item Then, we aim to jointly design the beamformer and the reflection of IRS to maximize the average sum-rate over all subcarriers. Based on the equivalence between sum-rate maximization and mean square error (MSE) minimization, the problem is converted to a multi-block/variable optimization, which can be solved by the classical block coordinate descent (BCD) method.
\item Finally, we evaluate our proposed design. We analyze the complexity and illustrate the convergence. Moreover, the performance of the proposed algorithm is validated by extensive simulation studies, which confirm the effectiveness of the design with the practical model compared to that with the ideal one.
\end{itemize}

\textit{Notations}:
Boldface lower-case and upper-case letters indicate column vectors and matrices, respectively.
$\mathbb{C}$ and $\mathbb{R}^{+}$ denote the set of complex and positive real numbers, respectively.
$(\cdot)^\ast$, $(\cdot)^T$, $(\cdot)^H$, and $(\cdot)^{-1}$ denote the conjugate, transpose, conjugate-transpose operations, and inversion, respectively.
$\mathbb{E} \{ \cdot \}$ represents statistical expectation.
$\Re \{ \cdot \}$ denotes the real part of a complex number.
$\mathbf{I}_L$ indicates an $L \times L$ identity matrix.
$\| \mathbf{A} \|_F$ denotes the Frobenius norm of matrix $\mathbf{A}$.
$\| \mathbf{a} \|_2$ denotes the $\ell_2$ norm of vector $\mathbf{a}$.
$\otimes$ denotes the Kronecker product.
$\mathrm{blkdiag}(\cdot)$ denotes a block matrix such that the main-diagonal blocks are matrices and all off-diagonal blocks are zero matrices.
Finally, $\mathbf{A}(i,:)$, $\mathbf{A}(:,j)$, and $\mathbf{A}(i,j)$ denote the $i$-th row, the $j$-th column, and the $(i,j)$-th element of matrix $\mathbf{A}$, respectively. $\mathbf{a}(i)$ denotes the $i$-th element of vector $\mathbf{a}$.

\section{Practical IRS Modeling}

\begin{figure}
\centering
\includegraphics[height=1.2 in]{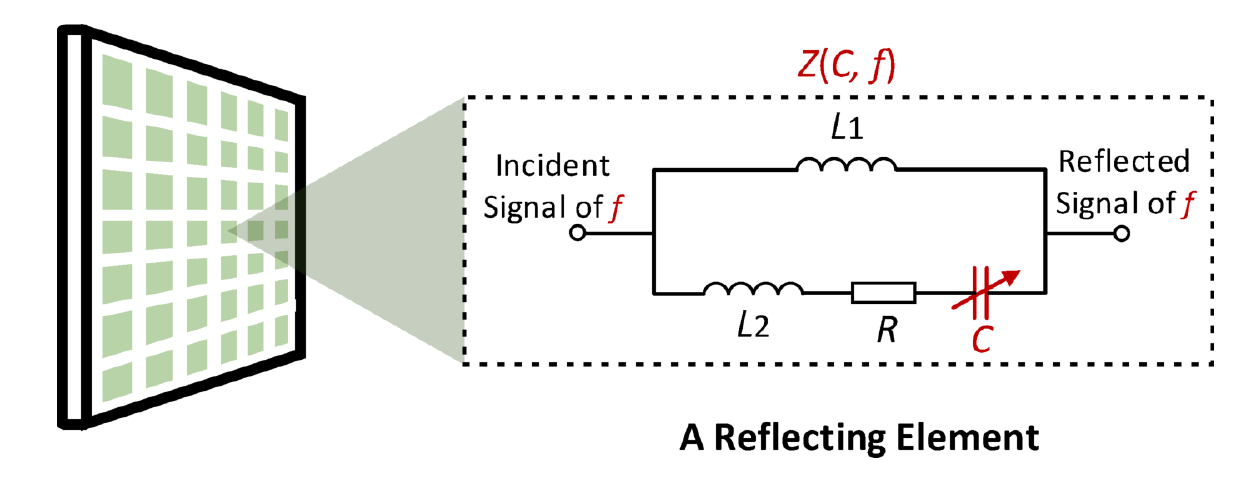}
\caption{The equivalent circuit of an IRS element.}\label{fig:IRS_element} \vspace{-0.0 cm}
\end{figure}

\begin{figure*}
\centering

\subfigure[]{
\begin{minipage}[t]{0.32\linewidth}
\centering
\includegraphics[width = 2.5 in]{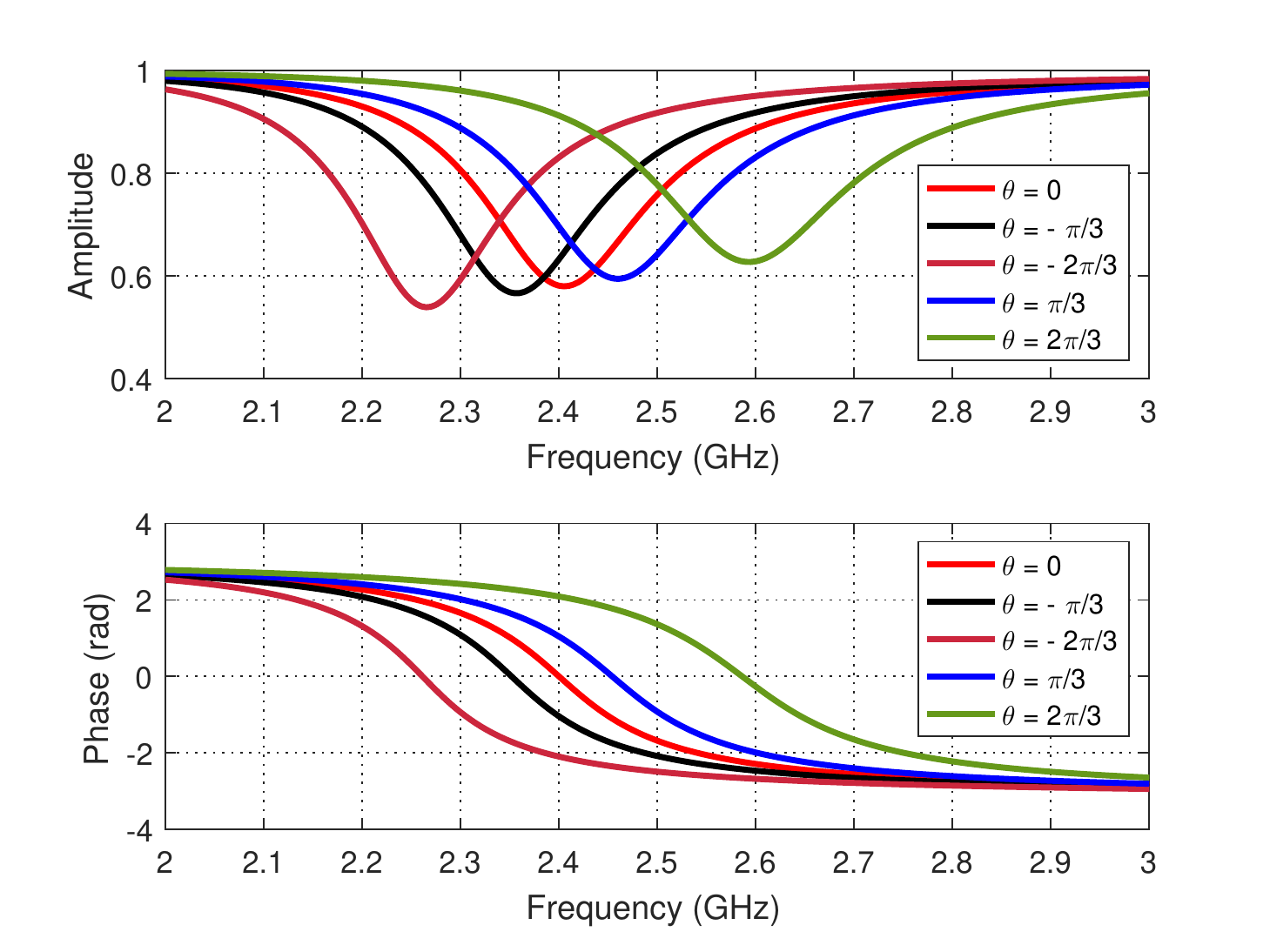}
\label{fig:IRS_model}
\end{minipage}%
}%
\subfigure[]{
\begin{minipage}[t]{0.32\linewidth}
\centering
\includegraphics[width = 2.5 in]{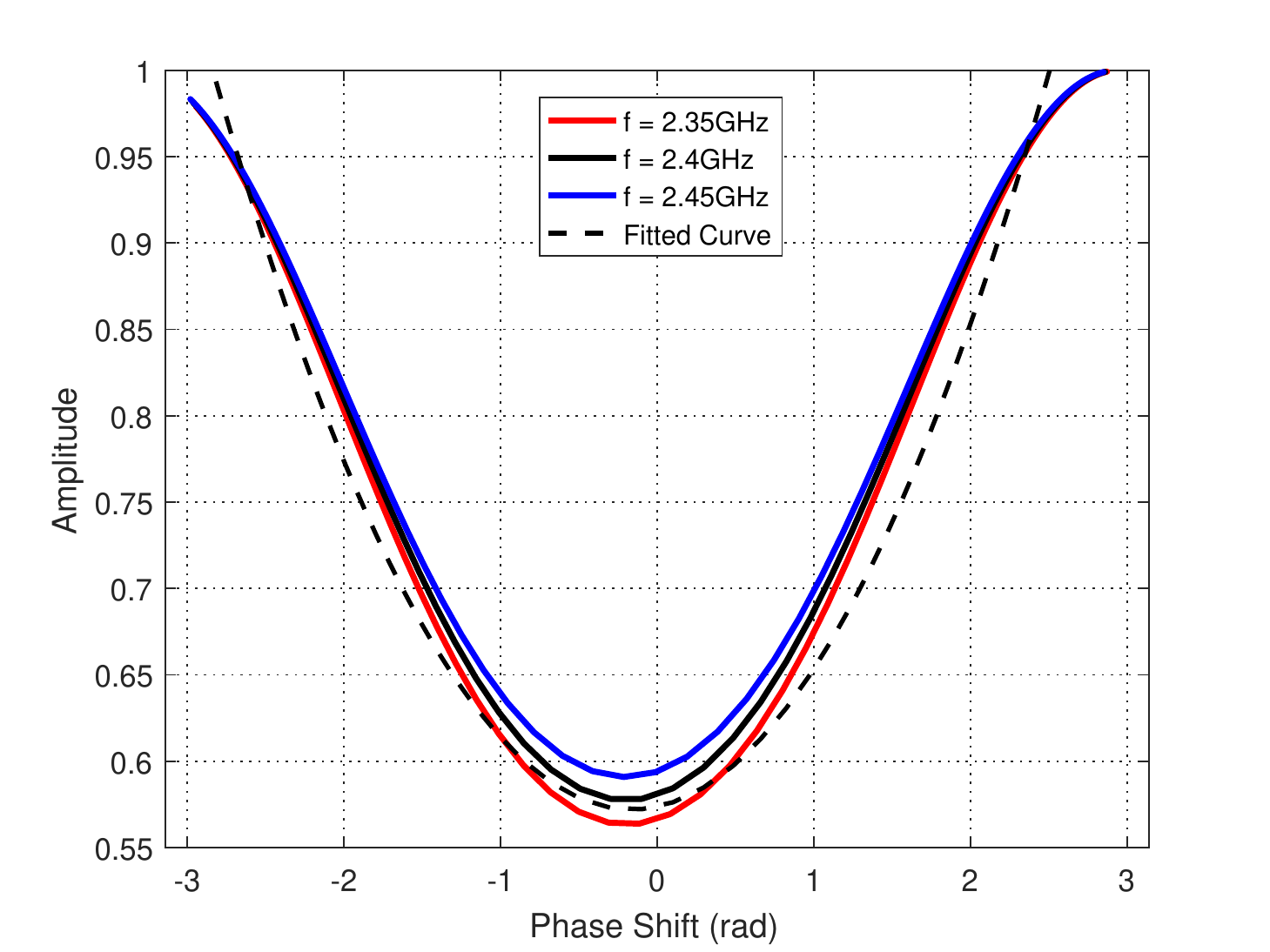}
\label{fig:amp_phs}
\end{minipage}%
}%
\subfigure[]{
\begin{minipage}[t]{0.32\linewidth}
\centering
\includegraphics[width = 2.5 in]{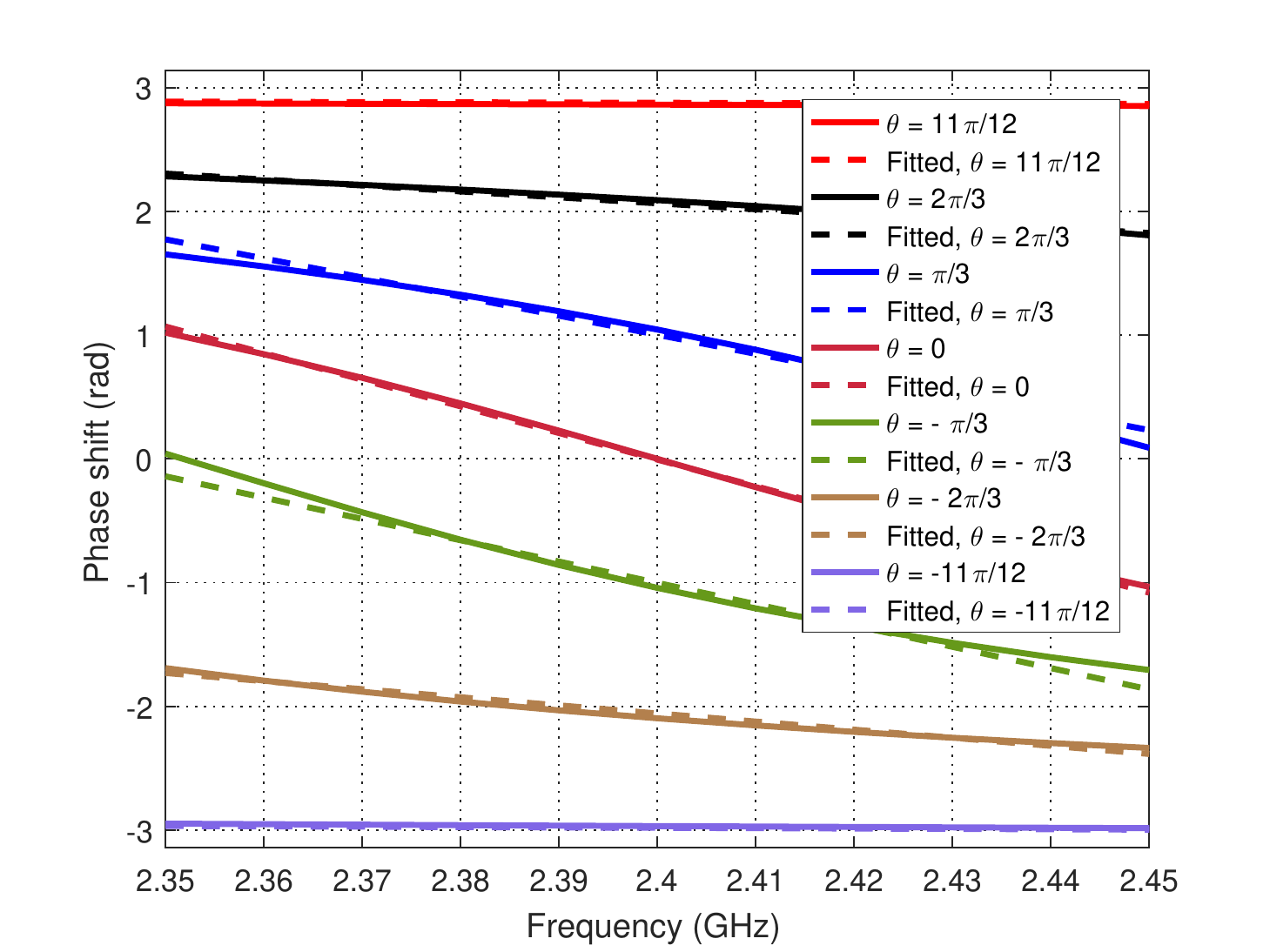}
\label{fig:phs_freq}
\end{minipage}%
}

\centering
\caption{(a) The illustration of the dual phase- and amplitude-squint \cite{W Cai}. With a certain phase shift $\theta$ for the carrier frequency $f_\mathrm{c} = 2.4$GHz, the amplitudes and phase shifts for other frequencies vary. With different phase shift $\theta$, the variation range and trend of the amplitudes and phase shifts for other frequencies also vary. (b) The relationship between the amplitude and the phase shift for corresponding frequencies. (c) The phase shift as a function of frequency.}
\vspace{0.1 cm}
\hrulefill
\end{figure*}

The hardware construction of IRS is usually based on the printed circuit board (PCB) with uniformly distributed reflecting elements on a planar surface.
A typical IRS generally consists of three layers: \textit{i)} An outer layer with a large number of metal elements printed on the PCB dielectric substrate; \textit{ii)} a copper plate to avoid the leakage of signal energy; \textit{iii)} a control circuit board for IRS control \cite{Q Wu 2019}.
A semiconductor device, such as the positive-intrinsic-negative (PIN) diode, is embedded into each metal element in the outer layer to tune the reflecting response, e.g., phase shift and amplitude.
The response of each reflecting element can be equivalently modeled as a parallel resonance circuit\footnote{Since the physical length of a reflecting element is usually smaller than the wavelength of the incident signal, the response of each reflecting element can usually be described by an equivalent lumped circuit model regardless of different types of realizations \cite{S Koziel}.} as shown in Fig. \ref{fig:IRS_element}. Thus, the impedance of an IRS element for the signal of frequency $f$ can be written as
\begin{equation}
Z(C,f) = \frac{j2\pi fL_1(j2\pi fL_2 + \frac{1}{j2\pi fC} + R)} {j2\pi fL_1 + j2\pi fL_2 + \frac{1}{j2\pi fC} + R},
\end{equation}
where $L_1$, $L_2$, $C$, and $R$ denote the metal plate inductance, outer layer inductance, effective capacitance, and the loss resistance, respectively.
The reflection coefficient of each IRS element, denoted as $\phi$, can be fundamentally modeled by the ratio of the power of the reflected signal to that of the incident one, which is therefore given by
\begin{equation}
\phi = \frac{Z(C,f) - Z_0}{Z(C,f) + Z_0},
\end{equation}
where $Z_0$ denotes the free space impedance.
Here, we should emphasize that the reflection of the IRS element is a function of $C$ and $f$.
When each element is controlled by selecting an appropriate capacitance $C$, the response of each element is also associated with the frequency of the incident signals.
Our previous work \cite{W Cai} has demonstrated that the same IRS element actually exhibits different responses (i.e., different amplitudes and phase shifts) to signals with different frequencies, which is referred to as \textit{dual phase- and amplitude-squint} effect in this paper.
Fig. \ref{fig:IRS_model} illustrates an example of the amplitude and phase shift variations of an IRS element as a function of frequency\footnote{In this paper, we adopt a typical setting following the device manual for a practical surface mount diode SMV1231-079 with equivalent parameters $L_1 = 2.5$ nH, $L_2 = 0.7$ nH, $R = 1 \Omega$, and the capacitance $C$ varying from 0.47 pF to 2.35 pF.}. We name the phase shift $\theta$ for signal of carrier frequency $f_\mathrm{c}$ as the basic phase shift (BPS) for clear and concise description. We can observe from Fig. \ref{fig:IRS_model} that, if we change the BPS $\theta$, the phase shifts and amplitudes for other frequencies will be quite different, which illustrates the severe beam deviations due to the dual phase- and amplitude-squint.
It is worth noting that it is an intrinsic phenomenon depending on the practical IRS circuit implementation, which cannot be simply ignored in realistic IRS-enhanced wideband systems.
Therefore, it is necessary to consider the dual phase- and amplitude-squint into account by developing an accurate reflection model of each IRS element, which is crucial for the following joint beamforming and reflecting design.

\begin{figure*}
\centering
\includegraphics[height=2.3in]{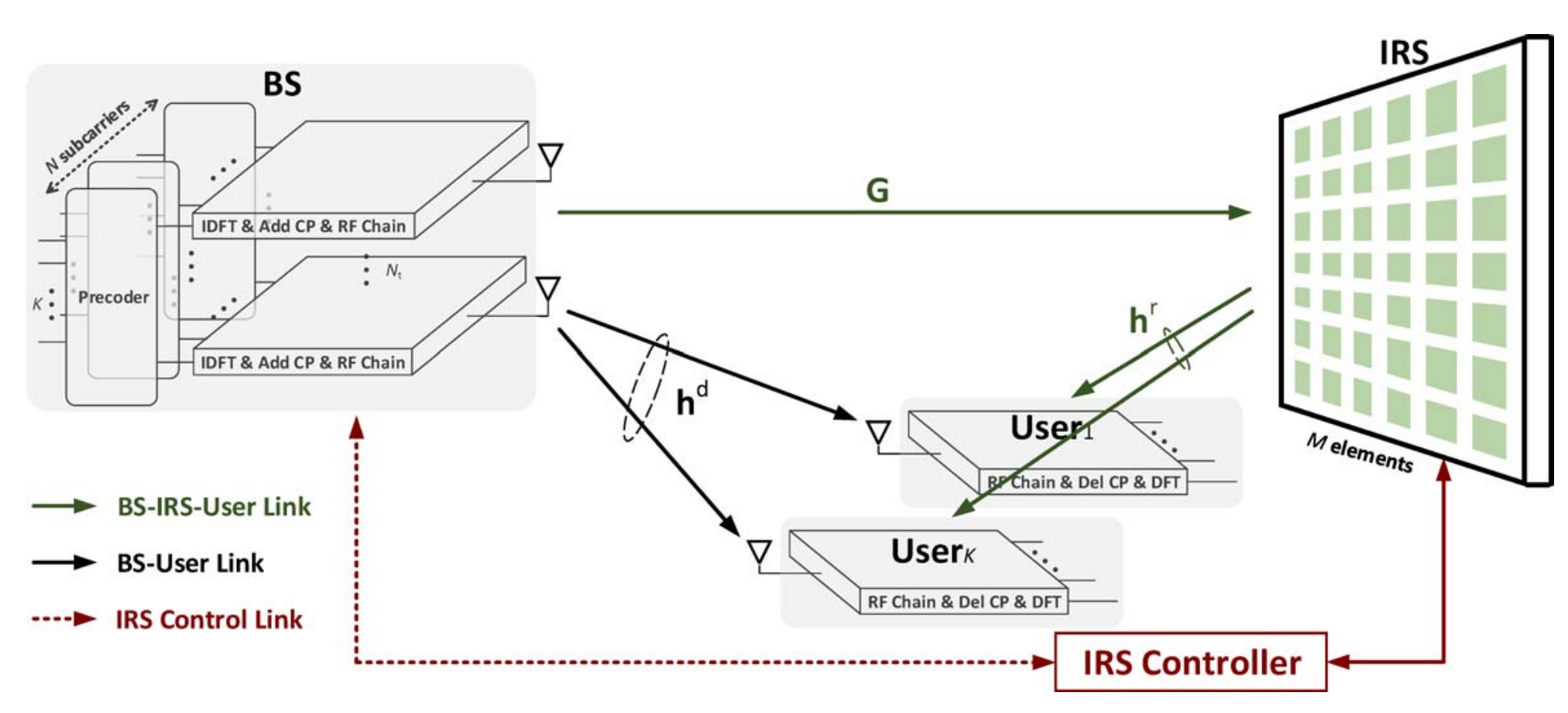}
\caption{The illustration of an IRS-enhanced MU-MISO-OFDM system.}\label{fig:LIS} \vspace{-0.0 cm}
\hrulefill
\end{figure*}

In \cite{W Cai}, we have established an accurate three-dimensional amplitude-phase-frequency model to describe the dual phase- and amplitude-squint.
Unfortunately, this model is so complicated that it may significantly increase the difficulty and complexity of IRS design.
To effectively simplify this model while maintaining its accuracy, we consider a more practical wideband situation that the relative bandwidth, i.e. the ratio of bandwidth and the carrier frequency $B/f_\mathrm{c}$, is less than $5\%$.
Take the case that the carrier frequency $f_\mathrm{c} = 2.4$ GHz and bandwidth $B = 100$ MHz as an example.
It can be observed from Fig. \ref{fig:amp_phs} that, the relationship between the amplitude of an IRS element and the phase shift for different frequencies can be viewed as a quadratic function. Moreover, within the bandwidth $B = 100$MHz, the curves for different frequencies do not have obvious difference, which motivates us to use a unified fit function for different frequencies.
Then in Fig. \ref{fig:phs_freq}, we can find that the phase shift as a function of frequency can be approximately fitted as a straight line.
When the BPS $\theta$ of one IRS element varies, the slope and intercept for the phase shift-frequency line will be different.

Motivated by the above findings, the simplified amplitude $\mathsf{F}(\theta,f)$ and phase shift $\mathsf{G}(\theta,f)$ of one certain IRS element corresponding to the incident signal of frequency $f$ can be modeled as
\begin{subequations}
\label{eq:IRS_model_simplified}
\begin{align}
&\mathsf{F}(\theta,f) = a_1\mathsf{G}^2(\theta,f) + b_1\mathsf{G}(\theta,f) + c_1, \\
&\mathsf{G}(\theta,f) = \mathsf{K}(\theta)f + \mathsf{B}(\theta), \\
&\mathsf{K}(\theta) = a_2\sin(b_2\theta + c_2) + a_3\sin(b_3\theta + c_3), \\
&\mathsf{B}(\theta) = a_4\sin(b_4\theta + c_4) + a_5\sin(b_5\theta + c_5),
\end{align}
\end{subequations}
where the functions $\mathsf{K}(\theta_m)$ and $\mathsf{B}(\theta_m)$ denote the slope and intercept for the phase shift-frequency line, respectively.
The central frequency of each subcarrier $f_i$ (GHz) is defined as
$f_i \triangleq f_\mathrm{c} + (i - \frac{N + 1}{2})\frac{B}{N}$, $\forall i \in \mathcal{N}$.
Parameters $\{a_i, b_i, c_i\}_{i=1}^5$ are related to specific circuit implementation. For practical examples showing in Fig. \ref{fig:IRS_model}, the values of these parameters are given in Table \ref{Tbale_1}. Specifically, the fitted results are shown as dash lines in Figs. \ref{fig:amp_phs} and \ref{fig:phs_freq}, which illustrate the accuracy of the proposed simplified model. In the next section, we attempt to utilize this practical model in an IRS-enhanced wideband MU-MISO-OFDM system and develop an effective algorithm to jointly design the transmit beamforming and IRS reflection.

\begin{center}
\begin{table}[t] \caption{Values of parameters for the simplified IRS model.}
\begin{center}
\begin{tabular}{cccccc}
\toprule
  ~ & $i=1$ & $i=2$ & $i=3$ & $i=4$ & $i=5$ \\
  \midrule
  $a_i$ &  0.06  & 11.27 & 10.88 & 89.64 & 26.11\\
  $b_i$ &  0.02  & 0.008996 & 0.9799 & 0.01268 & 0.9796\\
  $c_i$ &  0.5736 & -1.897 & -1.471 & 0.2899 & 1.673\\
 \bottomrule
\end{tabular}\label{Tbale_1}
\end{center} \vspace{-0.4 cm}
\end{table}
\end{center}

\section{System Model and Problem Formulation}

\subsection{System Model}

We consider a wideband MU-MISO-OFDM system with $N$ subcarriers, as shown in Fig. \ref{fig:LIS}.
The BS employs $N_\mathrm{t}$ antennas to communicate with $K$ single-antenna users.
This wireless transmission is assisted by an IRS of $M$ passive elements.
Denote $\mathcal{N} = \{1,\ldots,N\}$, $\mathcal{N}_\mathrm{t} = \{1,\ldots,N_\mathrm{t}\}$, $\mathcal{K} = \{1,\ldots,K\}$, and $\mathcal{M} = \{1,\ldots,M\}$ as the set of the indices of subcarriers, transmit antennas, users, and elements of the IRS, respectively.
The phase shifts of IRS elements are individually adjusted via a controller.
Next, we will describe the communication process in detail.

\textbf{Transmitter:} Let $\mathbf{s}_i \triangleq [s_{1,i}, \ldots, s_{K,i}]^T \in \mathbb{C}^{K}$ be the transmit symbols for all users associated with the $i$-th subcarrier with $\mathbb{E}\{\mathbf{s}_{i}\mathbf{s}_{i}^H\} = \mathbf{I}_{K}$, $\forall i\in \mathcal{N}$. The transmit symbol vector $\mathbf{s}_i$ is first digitally precoded by a precoder matrix $\mathbf{W}_i = [\mathbf{w}_{1,i}, \ldots, \mathbf{w}_{K,i}] \in \mathbb{C}^{N_\mathrm{t} \times K}$, $\mathbf{w}_{k,i} \in \mathbb{C}^{N_\mathrm{t}}$, $\forall i\in \mathcal{N}, \forall k \in \mathcal{K}$, in the frequency domain and then converted to the time domain by the inverse discrete Fourier transform (IDFT), which yields the overall time-domain signal $\widetilde{\mathbf{s}}$ as
\begin{equation}
\widetilde{\mathbf{s}} = (\mathbf{F}^{H}\otimes\mathbf{I}_{N_\mathrm{t}})\mathbf{W}\mathbf{s},
\label{eq:s_timedomain}
\end{equation}
where $\mathbf{F} \in \mathbb{C}^{N \times N}$ is the normalized discrete Fourier transform (DFT) matrix defined as $\mathbf{F}(m,n) \triangleq \frac{1}{\sqrt{N}}e^{\frac{-j2\pi (m-1)(n-1)}{N}}$, $\forall m,n \in \mathcal{N}$.
The overall precoding matrix $\mathbf{W}$ is given by $\mathbf{W} \triangleq \mathrm{blkdiag}(\mathbf{W}_1, \ldots, \mathbf{W}_N)$, and the overall transmit symbol vector $\mathbf{s}$ can be written as $\mathbf{s} \triangleq [\mathbf{s}_1^T, \ldots, \mathbf{s}_N^T]^T$.
After adding the cyclic prefix (CP) of size $N_\mathrm{cp}$, the signal is up-converted to the RF domain via $N_\mathrm{t}$ RF chains.

\textbf{Channel:}
In the considered wideband MU-MISO-OFDM system, the wideband channel\footnote{In this paper, exact and instantaneous channel state information (CSI) is assumed to be available at the BS. Existing works \cite{B Zheng}, \cite{Taha}-\cite{B Zheng OFDM CE} consider efficient channel estimation designs using ideal reflecting models. Recently, the authors in \cite{W. Yang} have focused on the channel estimation for practical IRS-enhanced OFDM system. Although it only considered the simplest single-input single-output (SISO) point-to-point scenario, it is easy to expand it to more general multiuser and/or MIMO systems.} from the BS to user $k$ is modeled by a $D$-tap ($D\le N_\mathrm{cp}$) finite-duration impulse response $\{\widetilde{\mathbf{h}}^{\mathrm{d}}_{k,0}, \ldots, \widetilde{\mathbf{h}}^{\mathrm{d}}_{k,D-1}\}$, where $\widetilde{\mathbf{h}}^{\mathrm{d}}_{k,d} \in \mathbb{C}^{N_\mathrm{t}}$, $d \in \mathcal{D} \triangleq \{0, \ldots, D-1\}$, $\forall k \in \mathcal{K}$, is the impulse response corresponding to the $d$-th delay tap\footnote{In this paper, we assume each impulse response follows the circularly symmetric complex Gaussion (CSCG) distribution.}.
Similarly, the wideband channel from the BS to the IRS is given by
$\{\widetilde{\mathbf{G}}_0, \ldots, \widetilde{\mathbf{G}}_{D-1}\}$ with $\widetilde{\mathbf{G}}_d \in \mathbb{C}^{M \times N_\mathrm{t}}$, $\forall d \in\mathcal{D}$.
The wideband channel from the IRS to user $k$ is given by
$\{\widetilde{\mathbf{h}}^{\mathrm{r}}_{k,0}, \ldots, \widetilde{\mathbf{h}}^{\mathrm{r}}_{k,D-1}\}$ with $\widetilde{\mathbf{h}}^{\mathrm{r}}_{k,d} \in \mathbb{C}^{M}, \forall d \in\mathcal{D}, \forall k \in \mathcal{K}$.

\begin{figure*}[t]
\begin{subequations}
\label{eq:freq_channel}
\begin{align}
\label{eq:fre_channel_a}
\mathbf{H}_k = &\mathbf{F}(\widetilde{\mathbf{H}}^\mathrm{d}_k + \widetilde{\mathbf{H}}^\mathrm{r}_k\mathbf{\Phi} \widetilde{\mathbf{G}})(\mathbf{F}^{H} \otimes\mathbf{I}_{N_\mathrm{t}})\\
\label{eq:fre_channel_b}
\overset{(\textrm{a})}= &\mathbf{F}(\widetilde{\mathbf{H}}^\mathrm{d}_k \mathbf{\Gamma}_1\mathbf{\Gamma}_1^T + \widetilde{\mathbf{H}}^\mathrm{r}_k
\mathbf{\Gamma}_2\mathbf{\Gamma}_2^T \mathbf{\Phi} \mathbf{\Gamma}_2\mathbf{\Gamma}_2^T \widetilde{\mathbf{G}}\mathbf{\Gamma}_1\mathbf{\Gamma}_1^T)
(\mathbf{F}^{H}\otimes\mathbf{I}_{N_\mathrm{t}}) \mathbf{\Gamma}_1\mathbf{\Gamma}_1^T\\
\label{eq:fre_channel_c}
= &\mathbf{F}(\underbrace{\widetilde{\mathbf{H}}^\mathrm{d}_k \mathbf{\Gamma}_1} + \underbrace{\widetilde{\mathbf{H}}^\mathrm{r}_k\mathbf{\Gamma}_2}\times \underbrace{\mathbf{\Gamma}_2^T\mathbf{\Phi} \mathbf{\Gamma}_2}\times\underbrace{\mathbf{\Gamma}_2^T \widetilde{\mathbf{G}}\mathbf{\Gamma}_1})\times
\underbrace{\mathbf{\Gamma}_1^T(\mathbf{F}^{H}\otimes\mathbf{I}_{N_\mathrm{t}}) \mathbf{\Gamma}_1}\times\mathbf{\Gamma}_1^T\\
\label{eq:fre_channel_d}
\overset{(\textrm{b})}= &\mathbf{F}([\widetilde{\mathbf{H}}^\mathrm{d}_{k,1}, \ldots, \widetilde{\mathbf{H}}^\mathrm{d}_{k,N_\mathrm{t}}] + [\widetilde{\mathbf{H}}^\mathrm{r}_{k,1}, \ldots, \widetilde{\mathbf{H}}^\mathrm{r}_{k,M}]\widetilde{\mathbf{\Phi}} \times
\left[\begin{array}{ccc}
     \widetilde{\mathbf{G}}_{1,1} &\ldots &\widetilde{\mathbf{G}}_{1,N_\mathrm{t}}\\
     \vdots &\ddots &\vdots\\
     \widetilde{\mathbf{G}}_{M,1} &\ldots &\widetilde{\mathbf{G}}_{M,N_\mathrm{t}}\\
  \end{array}\right])\times(\mathbf{I}_{N_\mathrm{t}} \otimes\mathbf{F}^{H})\mathbf{\Gamma}_1^T\\
\label{eq:fre_channel_e}
= &[\mathbf{F}\widetilde{\mathbf{H}}^\mathrm{d}_{k,1}\mathbf{F}^H + \mathbf{F}\sum\nolimits_{m=1}^M \widetilde{\mathbf{H}}^\mathrm{r}_{k,m}\widetilde{\mathbf{\Phi}}_m \widetilde{\mathbf{G}}_{m,1}\mathbf{F}^H, \ldots,
\mathbf{F}\widetilde{\mathbf{H}}^\mathrm{d}_{k,N_\mathrm{t}} \mathbf{F}^H + \mathbf{F}\sum\nolimits_{m=1}^M \widetilde{\mathbf{H}}^\mathrm{r}_{k,m}\widetilde{\mathbf{\Phi}}_m \widetilde{\mathbf{G}}_{m,N_\mathrm{t}}\mathbf{F}^H]\mathbf{\Gamma}_1^T\\
\label{eq:fre_channel_f}
\overset{(\textrm{c})}= &[\bm{\Lambda}^\mathrm{d}_{k,1} +\sum\nolimits_{m=1}^M\bm{\Lambda}^\mathrm{r}_{k,m} \widetilde{\mathbf{\Phi}}_m\bm{\Xi}_{m,1}, \ldots,
\bm{\Lambda}^\mathrm{d}_{k,N_\mathrm{t}} + \sum\nolimits_{m=1}^M\bm{\Lambda}^\mathrm{r}_{k,m} \widetilde{\mathbf{\Phi}}_m\bm{\Xi}_{m,N_\mathrm{t}}]\mathbf{\Gamma}_1^T\\
\label{eq:fre_channel_g}
\overset{(\textrm{d})} =
&\mathrm{diag}[(\mathbf{h}_{k,1}^{\mathrm{d}})^H + (\mathbf{h}_{k,1}^{\mathrm{r}})^H\mathbf{\Phi}_1\mathbf{G}_1, \ldots, (\mathbf{h}_{k,N}^{\mathrm{d}})^H + (\mathbf{h}_{k,N}^{\mathrm{r}})^H\mathbf{\Phi}_N\mathbf{G}_N], \forall k.
\end{align}
\end{subequations}
\hrulefill
\end{figure*}

\textbf{Receiver:}
After propagating through the wideband channels of both the BS-user link and the BS-IRS-user link, the overall time-domain signal $\widetilde{\mathbf{s}}$ is corrupted by additive Gaussion white noise (AGWN).
Down-converting to the baseband and removing the CP, we can obtain the time-domain received signal for all users. Specifically, we first define the block cyclic channel matrix $\widetilde{\mathbf{H}}^\mathrm{d}_k \in \mathbb{C}^{N \times NN_\mathrm{t}}$ of the BS-user link as
\begin{equation}
\non
\widetilde{\mathbf{H}}^\mathrm{d}_k = \left[
  \begin{array}{cccc}
     (\widetilde{\mathbf{h}}^{\mathrm{d}}_{k,0})^H &\mathbf{0}_{N_\mathrm{t}}^T &\ldots &(\widetilde{\mathbf{h}}^{\mathrm{d}}_{k,1})^H\\
     \vdots &(\widetilde{\mathbf{h}}^{\mathrm{d}}_{k,0})^H &\vdots &\vdots\\
     (\widetilde{\mathbf{h}}^{\mathrm{d}}_{k,D-1})^H&\vdots &\ddots &(\widetilde{\mathbf{h}}^{\mathrm{d}}_{k,D-1})^H\\
     \mathbf{0}_{N_\mathrm{t}}^T &(\widetilde{\mathbf{h}}^{\mathrm{d}}_{k,D-1})^H &\ddots &\vdots\\
     \vdots &\vdots &\vdots &\mathbf{0}_{N_\mathrm{t}}^T \\
     \mathbf{0}_{N_\mathrm{t}}^T &\mathbf{0}_{N_\mathrm{t}}^T &\ldots &(\widetilde{\mathbf{h}}^{\mathrm{d}}_{k,0})^H\\
  \end{array}
\right],
\end{equation}
$\forall k\in\mathcal{K}$.
Similarly, we define $[\widetilde{\mathbf{G}}^H_0, \ldots, \widetilde{\mathbf{G}}^H_{D-1}, \mathbf{0}_{N_\mathrm{t} \times M}$, $\ldots, \mathbf{0}_{N_\mathrm{t} \times M}]^H$ as the first block column of the block cyclic channel matrix $\widetilde{\mathbf{G}} \in \mathbb{C}^{MN \times NN_\mathrm{t}}$ of the BS-IRS link and $[\widetilde{\mathbf{h}}^{\mathrm{r}}_{k,0}, \ldots, \widetilde{\mathbf{h}}^{\mathrm{r}}_{k,D-1}, \mathbf{0}_{M}, \ldots, \mathbf{0}_{M}]^H$ as the first block column of the block cyclic channel matrix $\widetilde{\mathbf{H}}^\mathrm{r}_k \in \mathbb{C}^{N \times NM}$ of the IRS-user link. Then the time-domain received signal for user $k$ is given by
\begin{equation}
\widetilde{\mathbf{y}}_k = (\widetilde{\mathbf{H}}^\mathrm{d}_k + \widetilde{\mathbf{H}}^\mathrm{r}_k\mathbf{\Phi} \widetilde{\mathbf{G}})(\mathbf{F}^{H}\otimes\mathbf{I}_{N_\mathrm{t}}) \mathbf{W}\mathbf{s} + \widetilde{\mathbf{n}}_k, \forall k,
\label{eq:y_timedomain}
\end{equation}
where $\widetilde{\mathbf{n}}_k \in \mathcal{CN}(\mathbf{0}, \sigma^2\mathbf{I}_{N})$, $\forall k \in \mathcal{K}$, is the AGWN.
The reflection matrix $\mathbf{\Phi}$ of IRS is defined as $\mathbf{\Phi} = \mathrm{blkdiag}(\mathbf{\Phi}_1, \ldots, \mathbf{\Phi}_N)$, where $\mathbf{\Phi}_i \triangleq \mathrm{diag}(\phi_{i,1}, \ldots, \phi_{i,M})$, $\forall i \in \mathcal{N}$.
Here, $\phi_{i,m}$ denotes the reflection coefficient of the $m$-th IRS element for the $i$-th subcarrier.
Different from the ideal model in which each element exhibits the same reflection coefficient for different frequencies, i.e.,
\begin{equation}
|\phi_{i,m}| = 1, \angle\phi_{1,m} = \ldots = \angle\phi_{N,m}, \forall i, \forall m,
\end{equation}
we adopt the practical model presented in the previous section.
In particular, the reflection amplitude and phase shift of $\phi_{i,m}$ actually vary with the BPS $\theta_m$ and follow the relationship given in (\ref{eq:IRS_model_simplified}), i.e.,
\begin{equation}
|\phi_{i,m}| = \mathsf{F}(\theta_m,f_i), \angle\phi_{i,m} = \mathsf{G}(\theta_m,f_i), \forall i, \forall m.
\end{equation}
After operating DFT, the received signal in the frequency domain can be written as
\begin{equation}
\label{eq:yf_k}
\begin{aligned}
\mathbf{y}_k &= \mathbf{F}(\widetilde{\mathbf{H}}^\mathrm{d}_k + \widetilde{\mathbf{H}}^\mathrm{r}_k\mathbf{\Phi} \widetilde{\mathbf{G}})(\mathbf{F}^{H}\otimes\mathbf{I}_{N_\mathrm{t}}) \mathbf{W}\mathbf{s} + \mathbf{n}_k, \\
&= \mathbf{H}_k\mathbf{W}\mathbf{s} + \mathbf{n}_k, \forall k,
\end{aligned}
\end{equation}
where $\mathbf{n}_k \triangleq \mathbf{F}\widetilde{\mathbf{n}}_k, \forall k \in \mathcal{K}$.
The equivalent frequency-domain channel $\mathbf{H}_k$ for the $k$-th user is given by (\ref{eq:freq_channel}) on the top of this page, where the derivation of each step holds based on the following definitions and/or theorems:
\begin{enumerate}[(a):]
\item In equation (\ref{eq:fre_channel_b}), we introduce two column-wise permutation square matrices $\mathbf{\Gamma}_1$ and $\mathbf{\Gamma}_2$, $\mathbf{\Gamma}_1\mathbf{\Gamma}_1^T = \mathbf{I}_{NN_\mathrm{t}}, \mathbf{\Gamma}_2\mathbf{\Gamma}_2^T = \mathbf{I}_{NM}$, which convert a block cyclic matrix to several cyclic matrices arranged in rows \cite{SPAWC 2017}.
\item Let us define cyclic channel matrices for three links $\widetilde{\mathbf{H}}^\mathrm{d}_{k,n} \in \mathbb{C}^{N\times N}$, $\widetilde{\mathbf{G}}_{m,n} \in \mathbb{C}^{N\times N}$, and $\widetilde{\mathbf{H}}^\mathrm{r}_{k,m} \in \mathbb{C}^{N\times N}$, $\forall m \in \mathcal{M}, \forall n \in \mathcal{N}_\mathrm{t}, \forall k \in \mathcal{K}$, as
\begin{equation}
\begin{aligned}
&\widetilde{\mathbf{H}}^\mathrm{d}_{k,n}(:,i)\triangleq \widetilde{\mathbf{H}}^\mathrm{d}_{k}(:,n+(i-1)N_\mathrm{t}),\\
&\widetilde{\mathbf{G}}_{m,n}(p,q)\triangleq\widetilde{\mathbf{G}}(m+(p-1)M, n+(q-1)N_\mathrm{t}),\\
&\widetilde{\mathbf{H}}^\mathrm{r}_{k,m}(:,i)\triangleq \widetilde{\mathbf{H}}^\mathrm{r}_{k}(:,m + (i-1)M), ~\forall i, p, q,
\end{aligned}
\end{equation}
then the block cyclic channels can be arranged as a sequence of cyclic matrices, e.g., $\widetilde{\mathbf{H}}^\mathrm{d}_{k} \mathbf{\Gamma}_1 = [\widetilde{\mathbf{H}}^\mathrm{d}_{k,1}, \ldots, \widetilde{\mathbf{H}}^\mathrm{d}_{k,N_\mathrm{t}}]$, $\forall k \in \mathcal{K}$.
Similarly, by defining the rearranged reflection matrix $\widetilde{\mathbf{\Phi}} \triangleq \mathrm{blkdiag}(\widetilde{\mathbf{\Phi}}_1, \ldots, \widetilde{\mathbf{\Phi}}_M)$, $\widetilde{\mathbf{\Phi}}_m \triangleq \mathrm{diag}(\phi_{1,m}, \ldots, \phi_{N,m})$, $\forall m \in \mathcal{M}$, we can obtain $\mathbf{\Gamma}_2^T\mathbf{\Phi}\mathbf{\Gamma}_2 = \widetilde{\mathbf{\Phi}}$.
\item Since $\widetilde{\mathbf{H}}^\mathrm{d}_{k,n}$, $\widetilde{\mathbf{G}}_{m,n}$, as well as $\widetilde{\mathbf{H}}^\mathrm{r}_{k,m}$ in equation (\ref{eq:fre_channel_e}) are all cyclic matrices, they can be diagonalized by the DFT matrix. Specifically, we have $\mathbf{F}\widetilde{\mathbf{H}}^\mathrm{d}_{k,n}\mathbf{F}^H = \bm{\Lambda}^\mathrm{d}_{k,n}$, $\mathbf{F}\widetilde{\mathbf{H}}^\mathrm{r}_{k,m}\mathbf{F}^H = \bm{\Lambda}^\mathrm{r}_{k,m}$, and $\mathbf{F}\widetilde{\mathbf{G}}_{m,n}\mathbf{F}^H = \bm{\Xi}_{m,n}$, in equation (\ref{eq:fre_channel_f}), where $\bm{\Lambda}^\mathrm{d}_{k,n}, \bm{\Lambda}^\mathrm{r}_{k,m}$, and $\bm{\Xi}_{m,n}$ are diagonal matrices whose diagonal elements are the corresponding eigenvalues of $\widetilde{\mathbf{H}}^\mathrm{d}_{k,n}$, $\widetilde{\mathbf{H}}^\mathrm{r}_{k,m}$, and $\widetilde{\mathbf{G}}_{m,n}$, respectively.
\item Finally, by re-arranging equation (\ref{eq:fre_channel_f}), we can obtain the frequency-domain channels $\mathbf{h}^\mathrm{d}_{k,i} \in \mathbb{C}^{N_\mathrm{t}}$,
$\mathbf{h}^\mathrm{r}_{k,i} \in \mathbb{C}^{M}$, and
$\mathbf{G}_i \in \mathbb{C}^{M \times N_\mathrm{t}}$, $\forall k \in \mathcal{K}$, $\forall i \in \mathcal{N}$, which are defined as
\begin{equation}
\begin{aligned}
&\mathbf{h}^\mathrm{d}_{k,i}(n) \triangleq (\bm{\Lambda}^\mathrm{d}_{k,n}(i,i))^*, ~~
\mathbf{h}^\mathrm{r}_{k,i}(m) \triangleq (\bm{\Lambda}^\mathrm{r}_{k,m}(i,i))^*, \\
&\mathbf{G}_i(m,n) \triangleq \bm{\Xi}_{m,n}(i,i), ~~\forall m \in \mathcal{M}, \forall n \in \mathcal{N}_\mathrm{t}.
\end{aligned}
\end{equation}
\end{enumerate}
Substituting (\ref{eq:fre_channel_g}) into (\ref{eq:yf_k}), we can obtain the received signal on the $i$-th subcarrier for user $k$ as
\begin{equation}
\begin{aligned}
y_{k,i}=& [(\mathbf{h}_{k,i}^{\mathrm{d}})^H + (\mathbf{h}_{k,i}^{\mathrm{r}})^H\mathbf{\Phi}_i\mathbf{G}_i] \mathbf{W}_i\mathbf{s}_i + n_{k,i}\\
=& [(\mathbf{h}_{k,i}^{\mathrm{d}})^H + (\mathbf{h}_{k,i}^{\mathrm{r}})^H\mathbf{\Phi}_i\mathbf{G}_i] \mathbf{w}_{k,i}s_{k,i} + [(\mathbf{h}_{k,i}^{\mathrm{d}})^H+\\
&(\mathbf{h}_{k,i}^{\mathrm{r}})^H\mathbf{\Phi}_i\mathbf{G}_i] \sum_{p=1,p\ne k}^K\mathbf{w}_{p,i}s_{p,i} + n_{k,i}, \forall k, \forall i,
\end{aligned} \label{eq:received signal}
\end{equation}
where $n_{k,i}$ denotes the $i$-th element of $\mathbf{n}_{k}$.
With the signal model for a MU-MISO-OFDM system, we will consider the corresponding joint beamforming and reflection design problem in the next section.

\section{Joint Transmit Beamformer and IRS Reflection design}

\subsection{Problem Formulation}

We begin with establishing the optimization problem. With the received signal given in (\ref{eq:received signal}), the signal-to-interference-plus-noise ratio (SINR) on the $i$-th subcarrier for user $k$ can be calculated as
\begin{equation}
\gamma_{k,i} = \frac{|[(\mathbf{h}_{k,i}^{\mathrm{d}})^H + (\mathbf{h}_{k,i}^{\mathrm{r}})^H\mathbf{\Phi}_i\mathbf{G}_i] \mathbf{w}_{k,i}|^2}{\sum_{p\ne k}|[(\mathbf{h}_{k,i}^{\mathrm{d}})^H+
(\mathbf{h}_{k,i}^{\mathrm{r}})^H\mathbf{\Phi}_i\mathbf{G}_i] \mathbf{w}_{p,i}|^2 + \sigma^2}, \forall k, \forall i.
\end{equation}

In this paper, our goal is to jointly design the transmit beamformer $\mathbf{W}$ and the BPS matrix $\mathbf{\Theta} \triangleq \mathrm{diag}(\theta_1, \ldots, \theta_M)$, which essentially control the IRS reflection of wideband signals, to maximize the average sum-rate for the MU-MISO-OFDM system, subject to the constraints of the phase shift matrix and the transmit power. Therefore, the joint transmit beamformer and IRS reflection design problem can be formulated as
\begin{subequations}
\label{eq:problem0}
\begin{align}
\max_{\mathbf{W}, \mathbf{\Theta}} ~~ &\frac{1}{N}\sum_{i=1}^N\sum_{k=1}^K \log_2(1 + \gamma_{k,i})\\
\label{eq:p0_b}
\mathrm{s.t.} ~~~~ &|\phi_{i,m}| = \mathsf{F}(\theta_m,f_i), \forall i,m,\\
\label{eq:p0_c}
&\angle\phi_{i,m} = \mathsf{G}(\theta_m,f_i), \forall i,m,\\
\label{eq:p0_d}
&\theta_m \in [-\pi, \pi], \forall m,\\
\label{eq:p0_e}
&\sum_{i=1}^N\|\mathbf{W}_i\|_F^2 \le P,
\end{align}
\end{subequations}
where $P$ is the total transmit power at the BS.

Problem (\ref{eq:problem0}) is difficult to solve due to the complex form of the objective and the non-convex constraints of the BPS matrix. Furthermore, it is worth noting that the amplitude and phase shift of each IRS element will change with different frequencies when considering practical IRS responses for wideband signals.
In other words, we focus on the design of BPS matrix $\mathbf{\Theta}$, but the response of practical IRS for signals with different subcarriers varies, i.e., reflection matrix $\mathbf{\Phi}_i$, $\forall i
\in \mathcal{N}$, are different at each subcarrier.
This fact will further complicate the problem.
To deal with these issues, in the next section, we attempt to first transform problem (\ref{eq:problem0}) into a more tractable multi-variable/block optimization and then iteratively cope with each block.

\subsection{Problem Reformulation}
To tackle the difficulty rising from the $\sum\log(\cdot)$ function and the fractional form of ``SINRs'' in problem (\ref{eq:problem0}), we first reformulate the original sum-rate maximization problem as a modified MSE minimization problem \cite{Q Shi 2011}.
Let us first define the modified MSE function for user $k$ on the $i$-th subcarrier as
\begin{equation}
\begin{aligned}
\mathsf{MSE}_{k,i} =
&\mathbb{E}\{(\varpi_{k,i}^\ast y_{k,i} - s_{k,i})(\varpi_{k,i}^\ast y_{k,i} - s_{k,i})^\ast\}\\
=&\sum_{p=1}^K|\varpi_{k,i}^\ast[(\mathbf{h}_{k,i}^{\mathrm{d}})^H+
(\mathbf{h}_{k,i}^{\mathrm{r}})^H\mathbf{\Phi}_i\mathbf{G}_i]\mathbf{w}_{p,i}|^2\\
&-2\Re\{\varpi_{k,i}^\ast[(\mathbf{h}_{k,i}^{\mathrm{d}})^H+
(\mathbf{h}_{k,i}^{\mathrm{r}})^H\mathbf{\Phi}_i\mathbf{G}_i]\mathbf{w}_{k,i}\}\\
&+ |\varpi_{k,i}|^2\sigma^2+ 1, \forall k, \forall i,
\end{aligned}
\end{equation}
where $\varpi_{k,i} \in \mathbb{C}, \forall k \in \mathcal{K}, \forall i \in \mathcal{N},$ are auxiliary variables.
By introducing weighting parameters $\rho_{k,i} \in \mathbb{R}^+, \forall k \in \mathcal{K}, \forall i \in \mathcal{N}$, problem (\ref{eq:problem0}) can be equivalently transformed into the following form \cite{Q Shi 2011}:
\begin{subequations}
\label{eq:problem1}
\begin{align}
\label{eq:p1_a}
&\max_{\mathbf{W}, \Theta, \bm{\rho, \varpi}} ~~ \frac{1}{N}\sum_{i=1}^N\sum_{k=1}^K (\log_2\rho_{k,i} - \rho_{k,i}\mathsf{MSE}_{k,i} + 1)\\
&~~~\mathrm{s.t.} ~~~~\textrm{(\ref{eq:p0_b})-(\ref{eq:p0_e})},
\end{align}
\end{subequations}
where $\bm{\rho}$ and $\bm{\varpi}$ denote the sets of variables $\rho_{k,i}$ and $\varpi_{k,i}$, $\forall k \in \mathcal{K}, \forall i \in \mathcal{N}$, respectively.
Now, the newly formulated problem (\ref{eq:problem1}) is more tractable than the original problem after removing the complex fractional term (i.e. SINRs) from the $\log(\cdot)$ function.
In particular, problem (\ref{eq:problem1}) is a typical multi-variable/block problem, which can be solved using classical block coordinate descent (BCD) iterative algorithms \cite{Bertsekas 1999}.
In the following subsection, we will decompose problem (\ref{eq:problem1}) into four block optimizations and discuss the solution for each block in details.

\subsection{Block Update}
\subsubsection{Weighting parameter $\bm{\rho}$}
Fixing beamformers $\mathbf{W}_i, \forall i \in \mathcal{N}$, the BPS matrix $\mathbf{\Theta}$, and auxiliary variables $\varpi_{k,i}, \forall k \in \mathcal{K}, \forall i \in \mathcal{N}$, the sub-problem with respect to the weighting parameter $\rho_{k,i}$ is given by
\begin{equation}
\max_{\rho_{k,i}} ~~ \log_2\rho_{k,i} - \rho_{k,i}\mathsf{MSE}_{k,i}, \forall k, \forall i,
\label{eq:sub_rho}
\end{equation}
and the optimal solution can be easily obtained by checking the first-order optimality condition of problem (\ref{eq:sub_rho}), i.e.,
\begin{equation}
\rho_{k,i}^\star = \mathsf{MSE}_{k,i}^{-1} = 1 + \gamma_{k,i}, \forall k, \forall i.
\label{eq:opt_rho}
\end{equation}

\subsubsection{Auxiliary variable $\bm{\varpi}$}
When the beamformers $\mathbf{W}_i, \forall i \in \mathcal{N}$, the BPS matrix $\mathbf{\Theta}$, and weighting parameters $\rho_{k,i}, \forall k \in \mathcal{K}, \forall i \in \mathcal{N}$, are all fixed, the sub-problem with respect to the auxiliary variable $\varpi_{k,i}$ can be formulated as
\begin{equation}
\min_{\varpi_{k,i}} ~~ \rho_{k,i}\mathsf{MSE}_{k,i}, \forall k, \forall i,
\label{eq:sub_varpi}
\end{equation}
which is an unconstrained convex problem. Thus, problem (\ref{eq:sub_varpi}) can be solved by setting the partial derivative of the objective in (\ref{eq:sub_varpi}) with respect to $\varpi_{k,i}$ to zero, which yields the optimal value of $\varpi_{k,i}$ as
\begin{equation}
\varpi_{k,i}^\star = \frac{[(\mathbf{h}_{k,i}^{\mathrm{d}})^H + (\mathbf{h}_{k,i}^{\mathrm{r}})^H\mathbf{\Phi}_i\mathbf{G}_i] \mathbf{w}_{k,i}}{\sum_{p=1}^K|[(\mathbf{h}_{k,i}^{\mathrm{d}})^H+
(\mathbf{h}_{k,i}^{\mathrm{r}})^H\mathbf{\Phi}_i\mathbf{G}_i] \mathbf{w}_{p,i}|^2 + \sigma^2}, \forall k, \forall i.
\label{eq:opt_varpi}
\end{equation}

\subsubsection{Beamformer $\mathbf{W}$}
With weighting parameters $\rho_{k,i}$, auxiliary variables $\varpi_{k,i}, \forall k \in \mathcal{K}, \forall i \in \mathcal{N}$, and the BPS matrix $\mathbf{\Theta}$ given, the sub-problem with respect to the beamformer $\mathbf{W}_i, \forall i \in \mathcal{N}$, can be written as
\begin{subequations}
\label{eq:sub_w}
\begin{align}
\non
&\min_{\mathbf{W}}\frac{1}{N}\sum_{i=1}^N\sum_{k=1}^K \rho_{k,i}\Big(\sum_{p=1}^K|\varpi_{k,i}^\ast [(\mathbf{h}_{k,i}^{\mathrm{d}})^H+
(\mathbf{h}_{k,i}^{\mathrm{r}})^H\mathbf{\Phi}_i\mathbf{G}_i]
\mathbf{w}_{p,i}|^2\\
\label{eq:sub_w_a}
&~~~~~~-2\Re\{\varpi_{k,i}^\ast[(\mathbf{h}_{k,i}^{\mathrm{d}})^H+
(\mathbf{h}_{k,i}^{\mathrm{r}})^H\mathbf{\Phi}_i\mathbf{G}_i] \mathbf{w}_{k,i}\}\Big)\\
\label{eq:sub_w_c}
&~\mathrm{s.t.}~~\sum_{i=1}^N\|\mathbf{W}_i\|_F^2 \le P.
\end{align}
\end{subequations}
For convenience, we define the equivalent channel $\mathbf{h}_{k,i} \triangleq \big(\varpi_{k,i}^\ast((\mathbf{h}_{k,i}^{\mathrm{d}})^H+
(\mathbf{h}_{k,i}^{\mathrm{r}})^H\mathbf{\Phi}_i\mathbf{G}_i)\big)^H$, $\forall k \in \mathcal{K}, \forall i \in \mathcal{N}$. Then, problem (\ref{eq:sub_w}) can be concisely rewritten as
\begin{subequations}
\label{eq:sub_w1}
\begin{align}
\label{eq:sub_w1_b}
&\min_{\mathbf{W}}\frac{1}{N}\sum_{i=1}^N\sum_{k=1}^K \rho_{k,i} \Big(\sum_{p=1}^K|\mathbf{h}_{k,i}^H\mathbf{w}_{p,i}|^2 -2\Re\{\mathbf{h}_{k,i}^H\mathbf{w}_{k,i}\}\Big)\\
\overset{(\textrm{a})}=&\min_{\mathbf{W}}\frac{1}{N}\sum_{i=1}^N\sum_{k=1}^K \Big(\sum_{p=1}^K\rho_{p,i}|\mathbf{h}_{p,i}^H\mathbf{w}_{k,i}|^2 -2\rho_{k,i}\Re\{\mathbf{h}_{k,i}^H\mathbf{w}_{k,i}\}\Big)\\
\label{eq:sub_w1_c}
&~\mathrm{s.t.}~~\sum_{i=1}^N\|\mathbf{W}_i\|_F^2 \le P,
\end{align}
\end{subequations}
where (a) holds by changing the order of summations.
Since the objective and constraint of problem (\ref{eq:sub_w1}) are all convex, this problem can be optimally solved using the classic Lagrange multiplier optimization. To be specific, by introducing a multiplier $\mu \ge 0$ corresponding to the power constraint (\ref{eq:sub_w1_c}), problem (\ref{eq:sub_w1}) can be transformed into an unconstrained Lagrangian optimization:
\begin{subequations}
\begin{align}
\non
&\min_{\mathbf{W},\mu}~\sum_{i=1}^N\sum_{k=1}^K \Big(\sum_{p=1}^K\rho_{p,i}|\mathbf{h}_{p,i}^H\mathbf{w}_{k,i}|^2 -2\rho_{k,i}\Re\{\mathbf{h}_{k,i}^H\mathbf{w}_{k,i}\}\Big) \\
&~~~~~~+ \mu\left(\sum_{i=1}^N\|\mathbf{W}_i\|_F^2 - P\right)\\
\non
=&\min_{\mathbf{W},\mu}~\sum_{i=1}^N\sum_{k=1}^K \Big(\mathbf{w}_{k,i}^H\sum_{p=1}^K\rho_{p,i}\mathbf{h}_{p,i} \mathbf{h}_{p,i}^H\mathbf{w}_{k,i} \\ &~~~~~~-2\rho_{k,i}\Re\{\mathbf{h}_{k,i}^H\mathbf{w}_{k,i}\}
+ \mu\mathbf{w}_{k,i}^H\mathbf{w}_{k,i}\Big) - \mu P.
\end{align}
\end{subequations}
Similar to the solution of the previous two blocks, this unconstrained convex problem can be solved by checking the first-order optimality condition, which yields the optimal beamforming vector as
\begin{equation}
\mathbf{w}_{k,i}^\star = \Big(\sum_{p=1}^K\rho_{p,i}\mathbf{h}_{p,i}\mathbf{h}_{p,i}^H + \mu\mathbf{I}_{N_\mathrm{t}}\Big)^{-1} \rho_{k,i}\mathbf{h}_{k,i}, \forall k, \forall i,
\label{eq:opt_w}
\end{equation}
where the optimal multiplier $\mu$ is associated with the total power constraint and can be easily determined using a bisection search over the set $\mathcal{S}_{\mu} \triangleq \{\mu \ge 0 ~|~ \sum_{i=1}^{N}\|\mathbf{W}^\star_i\|_F^2 \le P\}$.

\subsubsection{BPS matrix $\mathbf{\Theta}$}
Given weighting parameters $\rho_{k,i}$, auxiliary variables $\varpi_{k,i}$, and beamfomers $\mathbf{W}_i, \forall i \in \mathcal{N}, \forall k \in \mathcal{K}$, the sub-problem with respect to the BPS matrix $\mathbf{\Theta}$ can be presented as
\begin{subequations}
\label{eq:sub_phi}
\begin{align}
\non
&\min_{\mathbf{\Theta}}\frac{1}{N}\sum_{i=1}^N\sum_{k=1}^K \rho_{k,i}\Big(\sum_{p=1}^K|\varpi_{k,i}^\ast [(\mathbf{h}_{k,i}^{\mathrm{d}})^H+
(\mathbf{h}_{k,i}^{\mathrm{r}})^H\mathbf{\Phi}_i\mathbf{G}_i]
\mathbf{w}_{p,i}|^2\\
\label{eq:sub_phi_a}
&~~~~~~-2\Re\{\varpi_{k,i}^\ast[(\mathbf{h}_{k,i}^{\mathrm{d}})^H+
(\mathbf{h}_{k,i}^{\mathrm{r}})^H\mathbf{\Phi}_i\mathbf{G}_i] \mathbf{w}_{k,i}\}\Big)\\
\label{eq:sub_phi_b}
&~\mathrm{s.t.}~~\textrm{(\ref{eq:p0_b})-(\ref{eq:p0_d})}.
\end{align}
\end{subequations}
By defining $\bm{\phi}_i \triangleq [\phi_{i,1}, \ldots, \phi_{i,M}]^T$, $\overline{h^\mathrm{d}}_{k,p,i} \triangleq (\mathbf{h}_{k,i}^{\mathrm{d}})^H\mathbf{w}_{p,i}$, and $\mathbf{v}_{k,p,i} \triangleq [(\mathbf{h}_{k,i}^{\mathrm{r}})^H\mathrm{diag} (\mathbf{G}_i\mathbf{w}_{p,i})]^H, \forall k,p\in\mathcal{K}, \forall i \in\mathcal{N}$, problem (\ref{eq:sub_phi}) can be concisely rearranged as
\begin{subequations}
\label{eq:sub_phi1}
\begin{align}
\non
&\min_{\mathbf{\Theta}}\frac{1}{N}\sum_{i=1}^N\sum_{k=1}^K \rho_{k,i}\Big(\sum_{p=1}^K|\varpi_{k,i}^\ast (\overline{h^\mathrm{d}}_{k,p,i} + \mathbf{v}_{k,p,i}^H\bm{\phi}_i)|^2\\
\label{eq:obj_phi_a}
&~~~~-2\Re\{\varpi_{k,i}^\ast(\overline{h^\mathrm{d}}_{k,k,i}+
\mathbf{v}_{k,k,i}^H\bm{\phi}_i)\}\Big)\\
\label{eq:obj_phi_b}
=&\min_{\mathbf{\Theta}}~\frac{1}{N}\sum_{i=1}^{N}(\bm{\phi}_i^H\mathbf{A}_i \bm{\phi}_i - 2\Re\{\bm{\phi}_i^H\mathbf{b}_i\}),\\
&~\mathrm{s.t.}~~\textrm{(\ref{eq:p0_b})-(\ref{eq:p0_d})},
\end{align}
\end{subequations}
where
\begin{subequations}
\label{eq:obj_phi1}
\begin{align}
\label{eq:A}
\mathbf{A}_i &\triangleq \sum_{k=1}^K \rho_{k,i}|\varpi_{k,i}|^2\sum_{p=1}^K\mathbf{v}_{k,p,i} \mathbf{v}_{k,p,i}^H, \forall i,\\
\label{eq:b}
\mathbf{b}_i &\triangleq \sum_{k=1}^K\rho_{k,i}\Big(\varpi_{k,i} \mathbf{v}_{k,k,i}-|\varpi_{k,i}|^2\sum_{p=1}^K\mathbf{v}_{k,p,i} \overline{h^\mathrm{d}}_{k,p,i}\Big), \forall i.
\end{align}
\end{subequations}

Problem (\ref{eq:sub_phi1}) is still difficult to solve since the BPS matrix $\mathbf{\Theta}$ to be optimized is embedded into a summation of $N$ complicated functions.
To simplify the design, one feasible solution is to decompose the joint optimization of the entire matrix $\mathbf{\Theta}$ into sub-problems, each of which deals with only one entry of $\mathbf{\Theta}$ while fixing others.
This alternative update of $\mathbf{\Theta}$ is conducted iteratively until the objective value converges.

Towards this end, we first split the objective (\ref{eq:obj_phi_b}) as
\begin{equation}
\begin{aligned}
&\frac{1}{N}\sum_{i=1}^N(\bm{\phi}_i^H\mathbf{A}_i\bm{\phi}_i - 2\Re\{\bm{\phi}_i^H\mathbf{b}_i\})\\
=&\frac{1}{N}\sum_{i=1}^N\sum_{m=1}^M\Big(\sum_{n=1}^M\mathbf{A}_i(m,n) \phi_{i,m}^\ast\phi_{i,n} - 2\Re\{\phi_{i,m}^\ast\mathbf{b}_i(m)\}\Big).
\end{aligned}
\end{equation}
Specifically, when we consider just one element while fixing other elements, e.g., the $m$-th element, the related objective is given by
\begin{equation}
\begin{aligned}
&\sum_{i=1}^N\Big(\sum_{n \ne m}(\mathbf{A}_i(m,n)\phi_{i,m}^\ast\phi_{i,n} + \mathbf{A}_i(n,m)\phi_{i,n}^\ast\phi_{i,m})\\
&~~~+ \mathbf{A}_i(m,m)|\phi_{i,m}|^2 - 2\Re\{\phi_{i,m}^\ast\mathbf{b}_i(m)\}\Big)\\
\overset{(\textrm{a})} = &\sum_{i=1}^N\Big(\sum_{n \ne m}(\mathbf{A}_i(m,n)\phi_{i,m}^\ast\phi_{i,n} + \mathbf{A}_i^*(m,n)\phi_{i,m}\phi_{i,n}^\ast)\\
&~~~+ \mathbf{A}_i(m,m)|\phi_{i,m}|^2 - 2\Re\{\phi_{i,m}^\ast\mathbf{b}_i(m)\}\Big)\\
= &\sum_{i=1}^N\Big(2\Re\Big\{\Big(\sum_{n \ne m}\mathbf{A}_i(m,n)\phi_{i,n}-\mathbf{b}_i(m)\Big)\phi_{i,m}^\ast\Big\} \\ &~~~+\mathbf{A}_i(m,m)|\phi_{i,m}|^2\Big),
\end{aligned}
\end{equation}
where (a) holds since $\mathbf{A}_i = \mathbf{A}_i^H, \forall i \in \mathcal{N}$. Then, the sub-problem with respect to the $m$-th BPS element $\theta_m$can be formulated as
\begin{subequations}
\label{eq:sub_theta1}
\begin{align}
\non
&\min_{\theta_m}~\sum_{i=1}^N\Big(2\Re\Big\{\Big(\sum_{n \ne m}\mathbf{A}_i(m,n)\phi_{i,n} - \mathbf{b}_i(m)\Big)\phi_{i,m}^\ast\Big\}\\
\label{eq:sub_theta_obj1}
&~~~~~~+ \mathbf{A}_i(m,m)|\theta_{i,m}|^2\Big)\\
&~\mathrm{s.t.}~~\textrm{(\ref{eq:p0_b})-(\ref{eq:p0_d})}.
\end{align}
\end{subequations}
We further define $\chi_{i,m} \triangleq \sum_{n \ne m}\mathbf{A}_i(m,n)\phi_{i,n} - \mathbf{b}_i(m)$, $\forall i \in\mathcal{N}, \forall m \in \mathcal{M}$, and substitute the constraints (\ref{eq:p0_b}), (\ref{eq:p0_c}) into the objective (\ref{eq:sub_theta_obj1}). Then, sub-problem (\ref{eq:sub_theta1}) can be reformulated as
\begin{subequations}
\label{eq:sub_theta2}
\begin{align}
\non
&\min_{\theta_m}~\sum_{i=1}^N\big(2|\chi_{i,m}|\mathsf{F}(\theta_m,f_i) \cos(\angle\chi_{i,m} - \mathsf{G}(\theta_m,f_i))\\
&~~~~~~+ \mathbf{A}_i(m,m)\mathsf{F}^2(\theta_m,f_i)\big)\\
&~\mathrm{s.t.}~~\theta_m \in [-\pi, \pi].
\end{align}
\end{subequations}

\begin{figure}[!t]
\centering

\subfigure[]{
\centering
\includegraphics[width = 3.2 in]{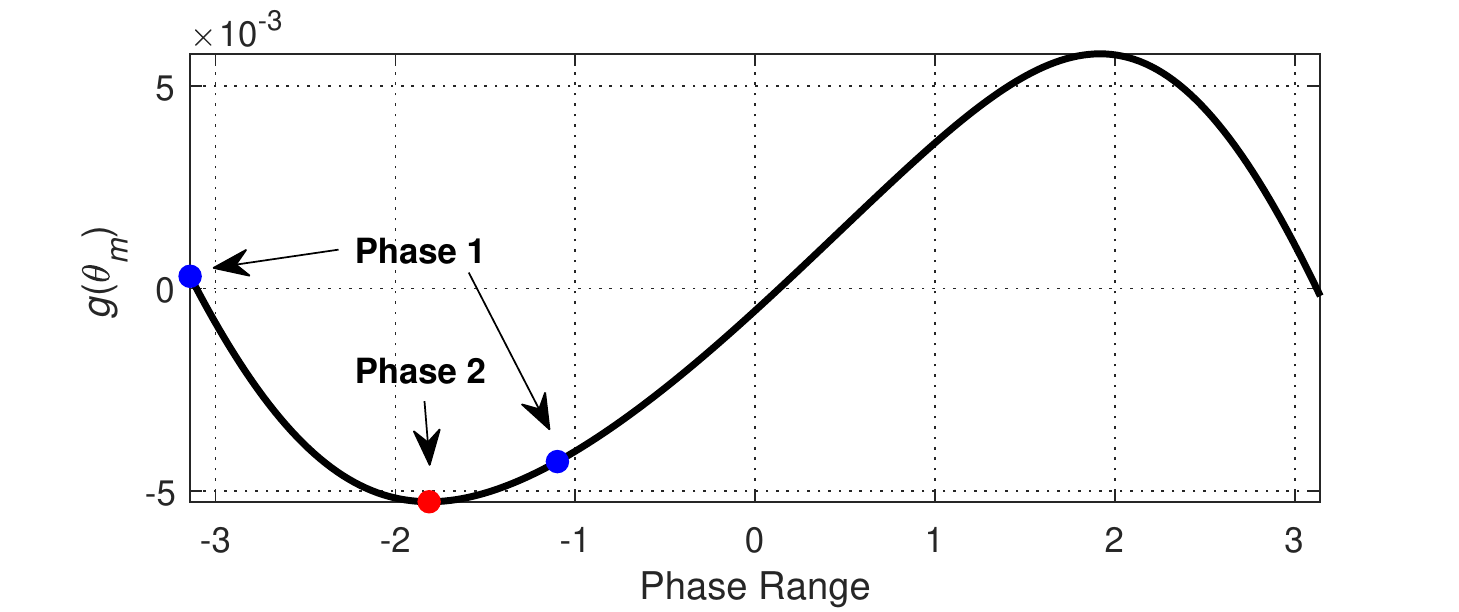}
\label{fig:ex1}
}%

\subfigure[]{
\centering
\includegraphics[width = 3.2 in]{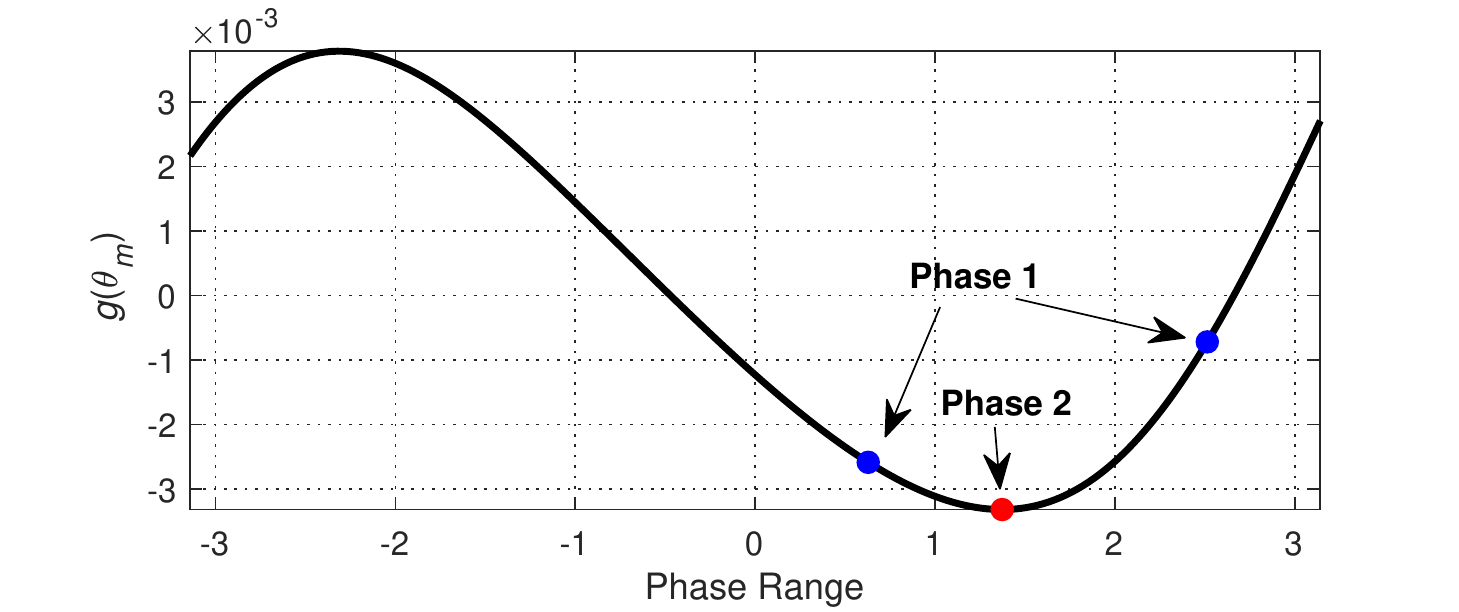}
\label{fig:ex2}
}%

\subfigure[]{
\centering
\includegraphics[width = 3.2 in]{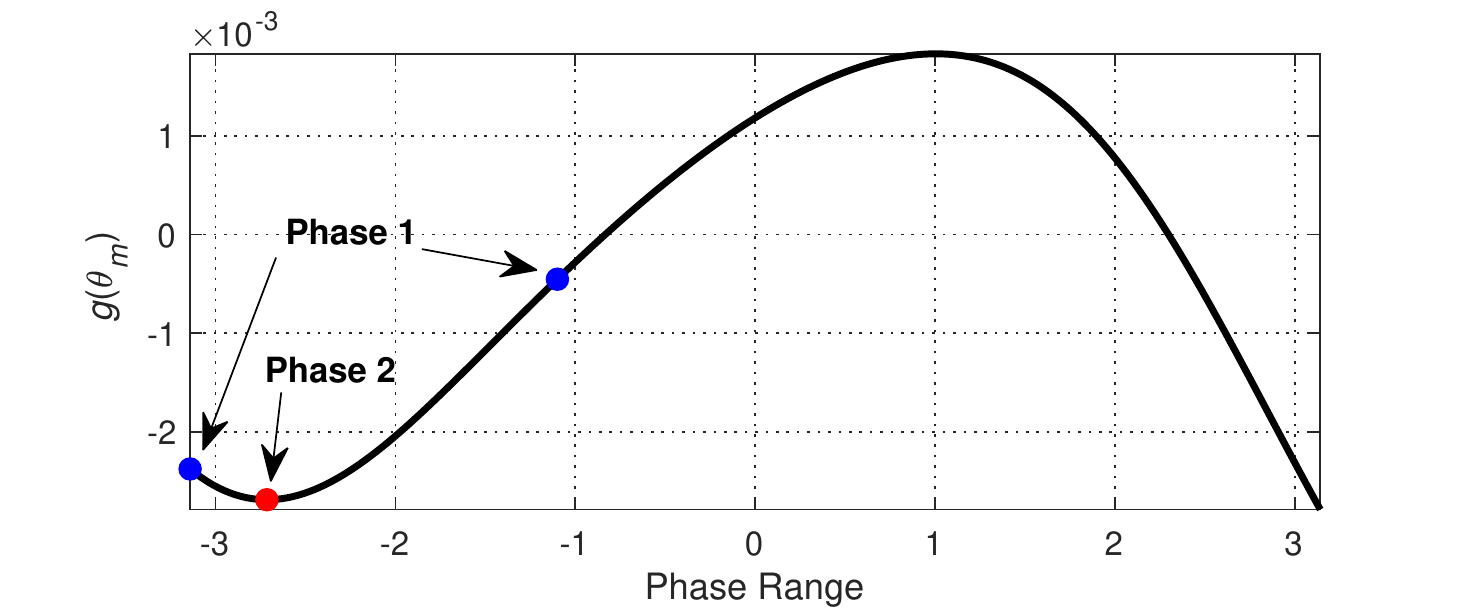}
\label{fig:ex3}}

\centering
\caption{Examples of the objective (\ref{eq:obj30}) as a function of the BPS within the range $[-\pi,\pi]$.}
\label{fig:obj29_ex}
\end{figure}

The objective of problem (\ref{eq:sub_theta2}) is a summation of $N$ complicated functions involving both trigonometric and quadratic terms, which is difficult to deal with. The computational complexity will be quite high when the numbers of IRS elements and/or subcarriers become large, which is the case for practical communication systems. To reduce the calculation complexity, we propose to further divide the whole bandwidth into $N_\mathrm{s}$ sub-bands, each of which comprises $S \triangleq N/N_\mathrm{s}$ subcarriers. By approximating each sub-band as a ``narrowband'' channel which has identical reflection coefficient configuration, problem (\ref{eq:sub_theta2}) can be further simplified as the optimization of a summation of much smaller number of functions, i.e.,
\begin{subequations}
\label{eq:sub_theta3}
\begin{align}
&\min_{\theta_m}~~~g(\theta_m)\\
&~\mathrm{s.t.}~~\theta_m \in [-\pi, \pi],
\end{align}
\end{subequations}
where the objective $g(\theta_m)$ is defined as
\begin{equation}
\begin{aligned}
\label{eq:obj30}
g(\theta_m) = &\sum_{i=1}^{N_\mathrm{s}}\big(2|\overline{\chi}_{i,m}|\mathsf{F}(\theta_m,f_{\mathrm{s},i}) \cos(\angle\overline{\chi}_{i,m} - \mathsf{G}(\theta_m,f_{\mathrm{s},i}))\\
&+ \overline{\alpha}_{i,m}\mathsf{F}^2(\theta_m,f_{\mathrm{s},i})\big),
\end{aligned}
\end{equation}
with $f_{\mathrm{s},i} \triangleq f_\mathrm{c} + (i - \frac{N_\mathrm{s} + 1}{2})\frac{B}{N_\mathrm{s}}$, $\overline{\chi}_{i,m} \triangleq \frac{1}{S} \sum_{j=1}^{S} \chi_{(i-1)S+j,m}$, and $\overline{\alpha}_{i,m} \triangleq \frac{1}{S} \sum_{j=1}^{S} \mathbf{A}_{(i-1)S+j}(m,m)$, $\forall i = 1,\ldots,N_\mathrm{s}$.
Unfortunately, the above problem is still difficult to solve since we cannot easily calculate the derivative of the objective and obtain the close-form solution.
To tackle this difficulty, we first try to explore the characteristic of the objective (\ref{eq:obj30}) with the aid of numerical experiments.
After numerous simulations (more than 5000 times), we find that objective (\ref{eq:obj30}) has only one minimum point within the range $[-\pi, \pi]$.
More concretely, objective (\ref{eq:obj30}) behaves like a kind of smooth double-peak-trough curve, whose minimum is achieved either at the minimum point or at two border points.
Some of examples are shown in Fig. \ref{fig:obj29_ex}.
Motivated by this finding, we propose a three-phase one-dimensional search method to efficiently find optimal solutions, which is summarized as follows:
\begin{enumerate}[\textit{Phase} 1:]
\item Narrow the search range by a success-failure method: Initialize a starting point $\theta_0$ as well as a step size $h > 0$. If $g(\theta_0 + h) < g(\theta_0)$, enlarge the step size and search forward until the objective rises; otherwise, search reversely until the objective rises.
\item Find the minimum point $\bar{\theta}$ by a golden section method: Successively section the search range which includes the minimum point in the golden ratio until reaching a predefined threshold.
\item Determine the minimum value: Compare the values of $g(\overline{\theta})$, $g(-\pi)$, and $g(\pi)$ to determine the minimal value as well as its corresponding phase shift.
\end{enumerate}
The details of the three-phase search algorithm are summarized in Algorithm 1. Furthermore, red points marked in Fig. \ref{fig:obj29_ex} are search results, which illustrate the accuracy of the proposed algorithm.

In realistic applications, the IRS is usually realized by finite- or even low-resolution phase shifters to effectively reduce the hardware consumption and cost. Therefore, we also consider the case that the BPS $\theta_m$ for IRS has discrete phases controlled by $b$ bits, which are uniformly spaced within the range $[-\pi,\pi)$, i.e.,
\begin{equation}
\theta_m \in \mathcal{F} \triangleq \{\frac{2\pi}{2^b}i - \pi | i = 0, 1, \ldots, 2^b\}, \forall m.
\label{eq:low_res}
\end{equation}
In this case, the IRS design sub-problem is given by
\begin{subequations}
\label{eq:sub_theta_discrete}
\begin{align}
\non
&\min_{\theta_m}~\sum_{i=1}^N\big(2|\chi_{i,m}|\mathsf{F}(\theta_m,f_i) \cos(\angle\chi_{i,m} - \mathsf{G}(\theta_m,f_i))\\
&~~~~~+ \mathbf{A}_i(m,m)\mathsf{F}^2(\theta_m,f_i)\big)\\
&~\mathrm{s.t.}~~\theta_m \in \mathcal{F}.
\end{align}
\end{subequations}
Similarly, we simplify this problem by dividing the whole bandwidth into several sub-bands, which yields the following problem:
\begin{subequations}
\label{eq:sub_theta_d1}
\begin{align}
&\min_{\theta_m}~~~g(\theta_m)\\
&~\mathrm{s.t.}~~\theta_m \in \mathcal{F}.
\end{align}
\end{subequations}
Thanks to the employment of low-resolution phase shifters, (i.e. $b \le 3$ bit) to realize the IRS, it is possible to perform an one-dimensional quick exhaustive search over the set $\mathcal{F}$ to find the optimal BPS element $\theta_{m}^\star$.

\begin{algorithm}[!t]
\caption{Three-Phase One-Dimensional Search}
\label{alg:AA1}
\begin{algorithmic}[1]
    \REQUIRE $f_{\mathrm{s},i}, \overline{\chi}_{i,m}, \overline{\alpha}_{i,m}, \forall i \in\mathcal{N}$.
    \ENSURE $\theta_m^\star$.
        \STATE {\textbf{Phase 1: Success-failure method}}
        \STATE {Initialize $\theta_0$, $h > 0$, $\theta_1 = \theta_0$, $\theta_2 = \theta_1 + h$.}
        \IF {$g(\theta_2) < g(\theta_1)$}
            \STATE {$\theta_3 = \theta_2 + h$}.
            \IF {$g(\theta_2) \le g(\theta_3)$}
                \STATE {Obtain the narrowed range $[\theta_\mathrm{l}, \theta_\mathrm{r}]$ as $\theta_\mathrm{l} = \min\{\theta_1, \theta_3\}$, $\theta_\mathrm{r} = \max\{\theta_1,\theta_3\}$, and stop.}
            \ELSE
                \STATE {$h = 2h$, $\theta_1 = \theta_2$, $\theta_2 = \theta_3$, $\theta_3 = \theta_2 + h$.}
                \STATE {Goto step 5.}
            \ENDIF
        \ELSE
            \STATE {$h = -h$, $\theta_3 = \theta_1$, $\theta_1 = \theta_2$, $\theta_2 = \theta_3$, $\theta_3 = \theta_2 + h$.}
            \STATE {Goto step 5.}
        \ENDIF
        \STATE {\textbf{Phase 2: Golden section method}}
        \STATE {Set $\overline{\theta}_\mathrm{l} = \theta_\mathrm{l} + 0.382(\theta_\mathrm{r} - \theta_\mathrm{l})$, $\overline{\theta}_\mathrm{r} = \theta_\mathrm{l} + 0.618(\theta_\mathrm{r} - \theta_\mathrm{l})$, $\epsilon$.}
        \WHILE {$\theta_\mathrm{r} - \theta_\mathrm{l} > \epsilon$}
            \IF {$g(\overline{\theta}_\mathrm{l}) \le g(\overline{\theta}_\mathrm{r})$}
                \STATE{$\theta_\mathrm{r} = \overline{\theta}_\mathrm{r}$, $\overline{\theta}_\mathrm{r} = \overline{\theta}_\mathrm{l}$, $\overline{\theta}_\mathrm{l} = \theta_\mathrm{l} + 0.382(\theta_\mathrm{r} - \theta_\mathrm{l})$.}
            \ELSE
                \STATE{$\theta_\mathrm{l} = \overline{\theta}_\mathrm{l}$, $\overline{\theta}_\mathrm{l} = \overline{\theta}_\mathrm{r}$, $\overline{\theta}_\mathrm{r} = \theta_\mathrm{l} + 0.618(\theta_\mathrm{r} - \theta_\mathrm{l})$.}
            \ENDIF
        \ENDWHILE
        \STATE{Obtain $\theta_m^\star = (\theta_\mathrm{l} + \theta_\mathrm{r})/2.$}
        \STATE{\textbf{Phase 3: Determine $\theta_m^\star$}}
        \IF {$g(\pi) \le g(\theta_m)$ and $g(\pi) \le g(-\pi)$}
            \STATE{$\theta_m^\star = \pi$.}
        \ELSIF {$g(-\pi) \le g(\theta_m)$ and $g(-\pi) \le g(\pi)$}
            \STATE{$\theta_m^\star = -\pi$.}
        \ENDIF
        \STATE {Return $\theta_m^\star$.}
\end{algorithmic}
\end{algorithm}

\subsubsection{Summary}
Having approaches to solve the above four sub-problems with respect to $\rho_{k,i}, \varpi_{k,i}, \mathbf{w}_{k,i}, \forall k \in \mathcal{K},\forall i \in \mathcal{N},$ and $\mathbf{\Theta}$, the overall procedure for the joint beamformer and IRS design is finally straightforward. Given appropriate initial values of $\mathbf{w}_{k,i}, \forall k \in \mathcal{K},\forall i \in \mathcal{N}$, and $\mathbf{\Theta}$, we iteratively update the above four blocks until convergence.
The proposed joint beamformer and IRS design algorithm is therefore summarized in Algorithm \ref{alg:AA2}.

\begin{algorithm}[!t]
\caption{Joint Transmit Beamformer and IRS Reflection Design}
\label{alg:AA2}
\begin{algorithmic}[1]
    \REQUIRE $\mathbf{h}_{k,i}^\mathrm{d},\mathbf{h}_{k,i}^\mathrm{r}, \mathbf{G}_{i}, \forall k \in \mathcal{K}, \forall i \in\mathcal{N}$, $P$, $B$.
    \ENSURE $\mathbf{w}_{k,i}^{\star}, \forall k \in \mathcal{K}, \forall i \in\mathcal{N}, \mathbf{\Theta}^\star$.
        \STATE {Initialize $\mathbf{\Theta}$, $\mathbf{w}_{k,i}, \forall k \in \mathcal{K},\forall i \in\mathcal{N}$.}
        \WHILE {no convergence of objective (\ref{eq:p1_a})}
            \STATE {Update $\rho_{k,i}, \forall k\in\mathcal{K},\forall i\in\mathcal{N}$ by (\ref{eq:opt_rho}).}
            \STATE {Update $\varpi_{k,i}, \forall k\in\mathcal{K},\forall i\in\mathcal{N}$ by (\ref{eq:opt_varpi}).}
            \STATE {Update $\mathbf{w}_{k,i}, \forall k\in\mathcal{K},\forall i\in\mathcal{N}$ by (\ref{eq:opt_w}).}
            \WHILE {no convergence of $\mathbf{\Theta}$}
                \FOR {$m=1:M$}
                    \STATE {Update $\theta_m$ by Algorithm 1 for continuous phases or by an exhaustive search for low-resolution phases.}
                \ENDFOR
            \ENDWHILE
        \ENDWHILE
        \STATE {Return $\mathbf{w}_{k,i}^{\star}, \forall k\in\mathcal{K},\forall i\in\mathcal{N}, \mathbf{\Theta}^\star$.}
\end{algorithmic}
\end{algorithm}

\subsection{Initialization}

For Algorithm \ref{alg:AA2}, an appropriate initialization for both the beamformer $\mathbf{W}$ and the BPS matrix $\mathbf{\Theta}$ is needed. Unfortunately, due to the complexity of the proposed practical IRS model in Sec. II, it is quite difficult to quickly and easily find a good initial value of $\mathbf{\Theta}$. Therefore, we just simply give random values of each BPS, i.e., $\theta_m, \forall m \in \mathcal{M}$, is uniformly selected within the range $[-\pi, \pi]$ (in the case of employing low-resolution phase shifters, each BPS is randomly selected within the discrete set $\mathcal{F}$).

Then, with this initial $\mathbf{\Theta}$ and corresponding reflection matrix $\mathbf{\Phi}$, we perform the typical MMSE beamformer as the initial value of transmit beamformer $\mathbf{W}$, which is given by
\begin{equation}
\overline{\mathbf{w}}_{k,i} = \mathbf{\Psi}_i^{-1}[(\mathbf{h}_{k,i}^{\mathrm{d}})^H+
(\mathbf{h}_{k,i}^{\mathrm{r}})^H\mathbf{\Phi}_i\mathbf{G}_i]^H, \forall k, \forall i,
\end{equation}
where $\mathbf{\Psi}_i \triangleq \sum_{k=1}^K[(\mathbf{h}_{k,i}^{\mathrm{d}})^H+
(\mathbf{h}_{k,i}^{\mathrm{r}})^H\mathbf{\Phi}_i\mathbf{G}_i]^H [(\mathbf{h}_{k,i}^{\mathrm{d}})^H+
(\mathbf{h}_{k,i}^{\mathrm{r}})^H\mathbf{\Phi}_i\mathbf{G}_i] + \sigma^2\mathbf{I}$, $\forall i \in \mathcal{N}$. To satisfy the transmit power constraint, the beamformer is further normalized by
\begin{equation}
\mathbf{w}_{k,i} = \frac{\sqrt{P}\overline{\mathbf{w}}_{k,i}}{\sqrt{\sum_{i=1}^{N} \sum_{k=1}^K\|\overline{\mathbf{w}}_{k,i}\|_2^2}}, \forall k, \forall i.
\end{equation}

\subsection{Complexity Analysis}

\begin{figure*}
\centering
\includegraphics[height=2.1 in]{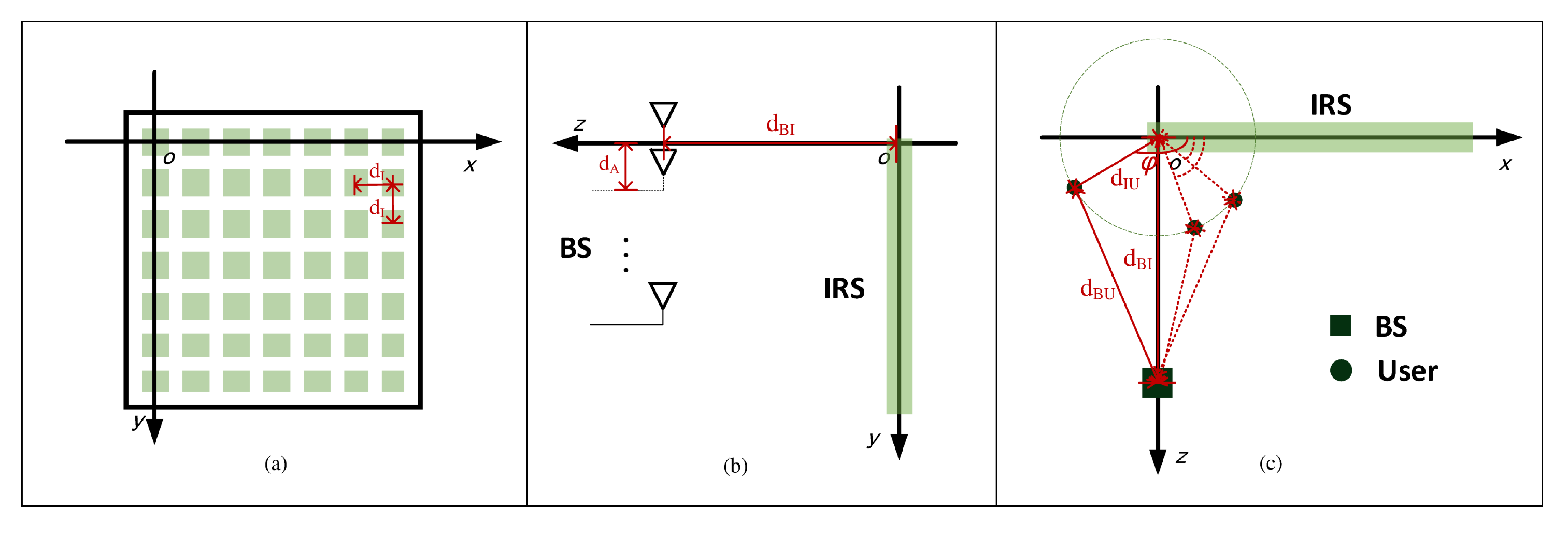}
\caption{An illustration of the relative position among the BS, IRS, and users.}\label{fig:distance} \vspace{-0.0 cm}
\hrulefill
\end{figure*}

In this subsection, we provide an analysis of the complexity for the proposed joint beamformer and IRS design algorithm. In each iteration, updating the weighting parameter $\bm \rho$ has a complexity of $\mathcal{O}(NK^2N_\mathrm{t}M^2)$ approximately; updating the auxiliary variable $\bm \varpi$ requires $\mathcal{O}(NK(K+1)N_\mathrm{t}M^2)$ operations; updating beamformer $\mathbf{W}$ requires about $\mathcal{O}(I_1NN_\mathrm{t}K(3M^2 + N_\mathrm{t}^2))$ operations, where the parameter $I_1$ denotes the iterations of bisection search. Finally, the order of complexity for updating BPS matrix $\mathbf{\Theta}$ for continuous phases is about $\mathcal{O}((5MN_\mathrm{t} + M^3)NK^2 + I_2N_\mathrm{s}M(I_3 + I_4))$, where $I_3$ and $I_4$ denotes the iterations for success-failure method and golden section method, respectively, and that for discrete phases is $\mathcal{O}((5MN_\mathrm{t} + M^3)NK^2 + I_5N_\mathrm{s}M2^b)$, where parameters $I_2$ and $I_5$ denote the numbers of iterations for calculating $\mathbf{\Theta}$. Therefore, the total complexity of the proposed algorithm is given by
\begin{subequations}
\begin{align}
\non
C_\mathrm{c} = &\mathcal{O}(I_\mathrm{c}( NK^2N_\mathrm{t}M^2 + NK(K+1)N_\mathrm{t}M^2\\
\non
&+ I_1NN_\mathrm{t}K(3M^2 + N_\mathrm{t}^2) + (5MN_\mathrm{t} + M^3)NK^2\\
&+ I_2N_\mathrm{s}M(I_3 + I_4)))\\
\overset{\textrm{(a)}}\approx &\mathcal{O}(I_\mathrm{c}(NK^2M^3 + 3I_1NN_\mathrm{t}KM^2 + I_2N_\mathrm{s}M(I_3 + I_4)))\\
\non
C_\mathrm{d} = &\mathcal{O}(I_\mathrm{c}( NK^2N_\mathrm{t}M^2 + NK(K+1)N_\mathrm{t}M^2\\
\non
&+ I_1NN_\mathrm{t}K(3M^2 + N_\mathrm{t}^2) + (5MN_\mathrm{t} + M^3)NK^2\\
&+ I_5N_\mathrm{s}M2^b))\\
\overset{\textrm{(a)}}\approx &\mathcal{O}(I_\mathrm{d}(NK^2M^3 + 3I_1NN_\mathrm{t}KM^2 + I_2N_\mathrm{s}MI_5N_\mathrm{s}M2^b)),
\end{align}
\end{subequations}
where $\mathrm{(a)}$ holds under assumptions $M \gg N_\mathrm{t}, M \gg K$. Parameters $I_\mathrm{c}$ (for continuous phases) and $I_\mathrm{d}$ (for discrete phases) are the numbers of iterations for Algorithm \ref{alg:AA2}. Simulation results in the next section show that, under different settings, the proposed algorithm for both continuous and discrete scenarios can converge within a small number of iterations, which demonstrates the efficiency of the proposed algorithm.

\section{Simulation Results}

\subsection{Simulation Settings}

In this section, we present simulation results to demonstrate the performance of the IRS-enhanced wideband MU-MISO-OFDM system by showing the average sum-rate of the proposed joint beamformer and IRS reflection design. In the considered IRS-enhanced MU-MISO-OFDM system, we assume the number of subcarriers is $N = 64$. The number of taps is set as $D = 16$ with half non-zero taps modeled as circularly symmetric complex Gaussian (CSCG) random variables.
The CP length is set to be $N_\mathrm{cp} = 16$.
The carrier frequency and bandwidth is given by $f_\mathrm{c} = 2.4$GHz and $B = 100$MHz, respectively.
The signal attenuation is set as $\zeta_0 = -30$ dB at a reference distance 1 m for all channels. The path loss exponent of the BS-IRS channel, the IRS-user channel, and the BS-user channel is set as $\varepsilon_\mathrm{BI} = 2.8$, $\varepsilon_\mathrm{IU} = 2.5$, and $\varepsilon_\mathrm{BU} = 3.7$, respectively.
The noise power at each user is set as $\sigma^2 = -70$ dBm.

In the following simulation results, we assume a three dimensional (3D) coordinate system is considered as shown in Fig. \ref{fig:distance}, where a uniform linear array (ULA) with antenna spacing $d_\mathrm{A} = 0.3$ m at the BS and a uniform planar array (UPA) with element-spacing $d_\mathrm{I} = 0.03$ m at the IRS and are located in y-z plane and x-y plane, respectively.
The distance between the reference antenna of the BS and the reference element of the IRS is given by $d_\mathrm{BI}$.
$K$ users are randomly located in x-z plane with the same distance $d_\mathrm{IU} = 1$ m as well as random phase $\varphi_k$ between the reference element of the IRS and the $k$-th user.
Based on the relative position given in Fig. \ref{fig:distance}, the distances between the $(p,q)$-th IRS element and the $k$-th user $d_\mathrm{IU}^{p,q,k}$,
the $n$-th antenna and the $k$-th user $d_\mathrm{BU}^{n,k}$, as well as the $n$-th antenna and the $(p,q)$-th IRS element $d_\mathrm{BI}^{n,p,q}$, are given by
\begin{equation}
\begin{aligned}
&d_\mathrm{IU}^{p,q,k}= \sqrt{(pd_\mathrm{I} - d_\mathrm{IU}\cos\varphi_k)^2 + q^2d_\mathrm{I}^2 + d_\mathrm{IU}^2\sin^2\varphi_k},\\
&d_\mathrm{BU}^{n,k}= \sqrt{(d_\mathrm{BI} - d_\mathrm{IU}\sin\varphi_k)^2 + n^2d_\mathrm{A}^2 + d_\mathrm{IU}^2\cos^2\varphi_k},\\
&d_\mathrm{BI}^{n,p,q}= \sqrt{(qd_\mathrm{I} - nd_\mathrm{A})^2 + p^2d_\mathrm{I}^2 + d_\mathrm{BI}^2}, \\
&\forall n \in \mathcal{N}_\mathrm{t}, \forall p,q = 1, \ldots, \sqrt{M}, \forall k \in \mathcal{K}.
\end{aligned}
\end{equation}
Then the fading component for the BS-IRS link, the BS-User link, and the IRS-User link is given by
\begin{equation}
\begin{aligned}
&\xi_\mathrm{BI}^{n,p,q} = \sqrt{\zeta_0 (d_\mathrm{BI}^{n,p,q})^{-\varepsilon_\mathrm{BI}}}, ~~ \xi_\mathrm{BU}^{n,k} = \sqrt{\zeta_0 (d_\mathrm{BU}^{n,k})^{-\varepsilon_\mathrm{BU}}}, \\
&\xi_\mathrm{IU}^{p,q,k} = \sqrt{\zeta_0 (d_\mathrm{IU}^{p,q,k})^{-\varepsilon_\mathrm{IU}}}, ~~~~~~\forall n,\forall p,q, \forall k.
\end{aligned}
\end{equation}
Thus, the channels for three links are given by
\begin{equation}
\begin{aligned}
&\widehat{\mathbf{h}}_{k,i}^\mathrm{r}(m) = \xi_\mathrm{IU}^{p,q,k}\mathbf{h}_{k,i}^\mathrm{r}(m), ~~~~
\widehat{\mathbf{h}}_{k,i}^\mathrm{d}(n) = \xi_\mathrm{BU}^{n,k}\mathbf{h}_{k,i}^\mathrm{d}(n),\\
&\widehat{\mathbf{G}}_i(m,n) = \xi_\mathrm{BI}^{n,p,q}\mathbf{G}_i(m,n),\\
&\forall n, \forall p,q, \forall k, \forall i \in \mathcal{N}, \forall m = (p-1)\sqrt{M}+q.
\end{aligned}
\end{equation}

\subsection{System Performance}

\begin{figure}[!t]
\centering
\subfigure[$N_\mathrm{t} = 4$, $M = 64$]{
{\label{fig:asr_vs_iter1}}
\includegraphics[height=2.5 in]{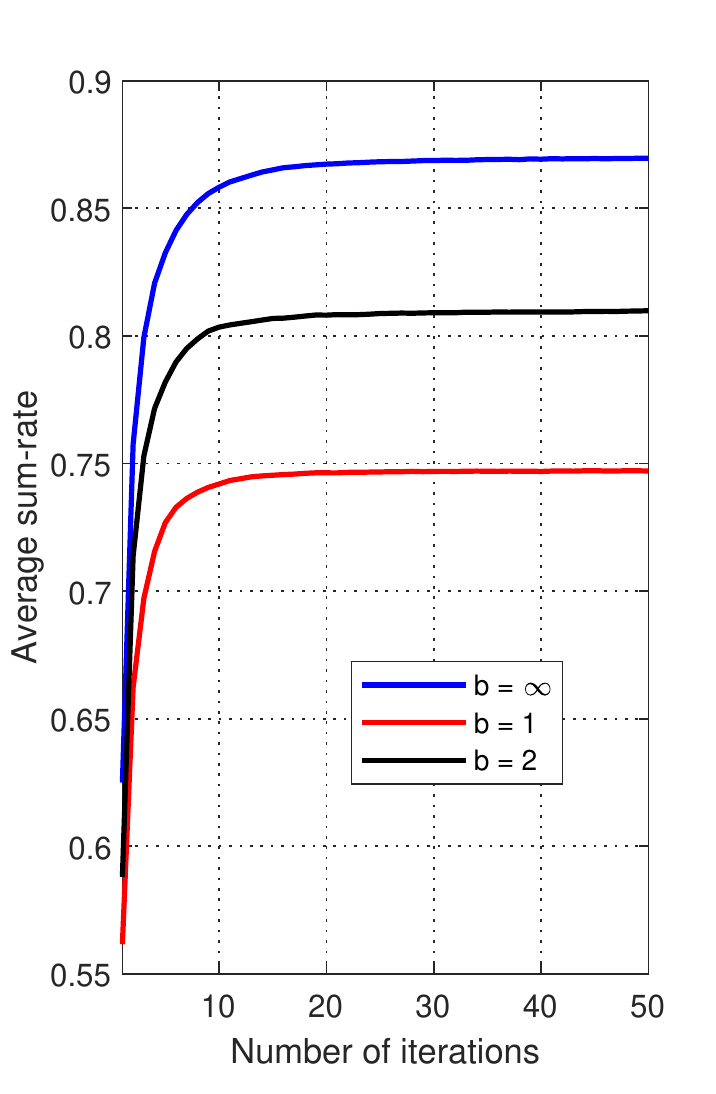}}
\hspace{-0.5cm}
\subfigure[$N_\mathrm{t} = 8$, $M = 144$]{
{\label{fig:asr_vs_iter2}}
\includegraphics[height=2.5 in]{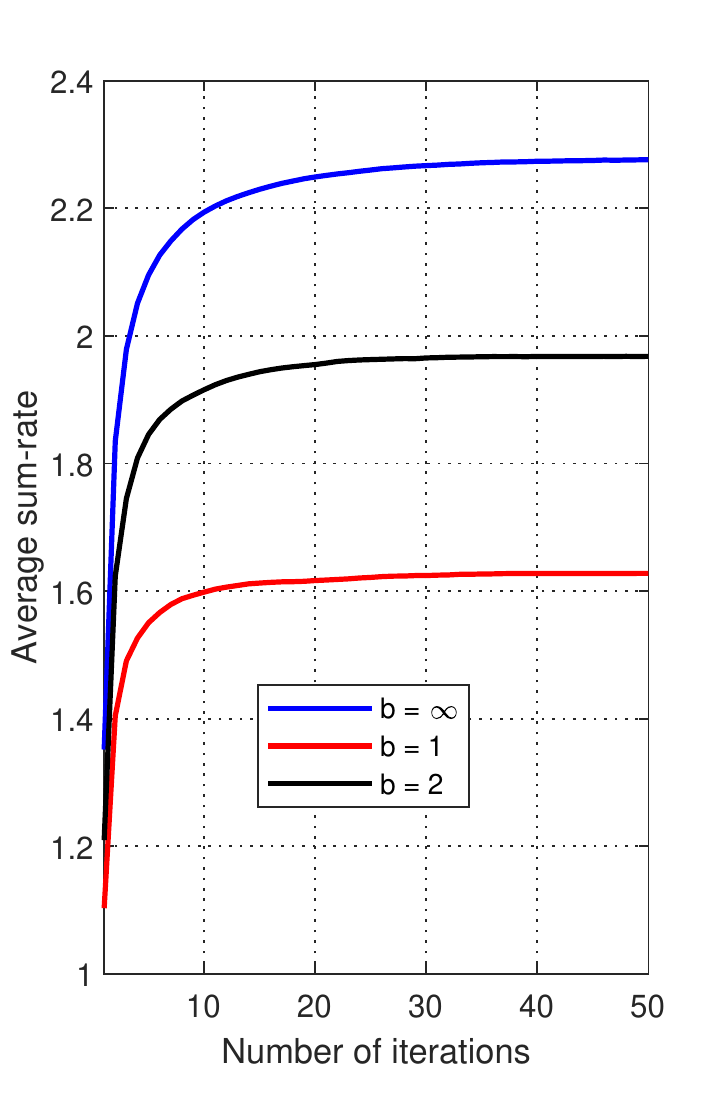}}
\caption{Average sum-rate versus the number of iterations ($K = 3$, $N = 64$, $N_\mathrm{s} = 4$, $P = -5$ dB).}
\label{fig:asr_vs_iter}
\end{figure}

\begin{figure}[!t]
\centering
\includegraphics[width=3.5 in]{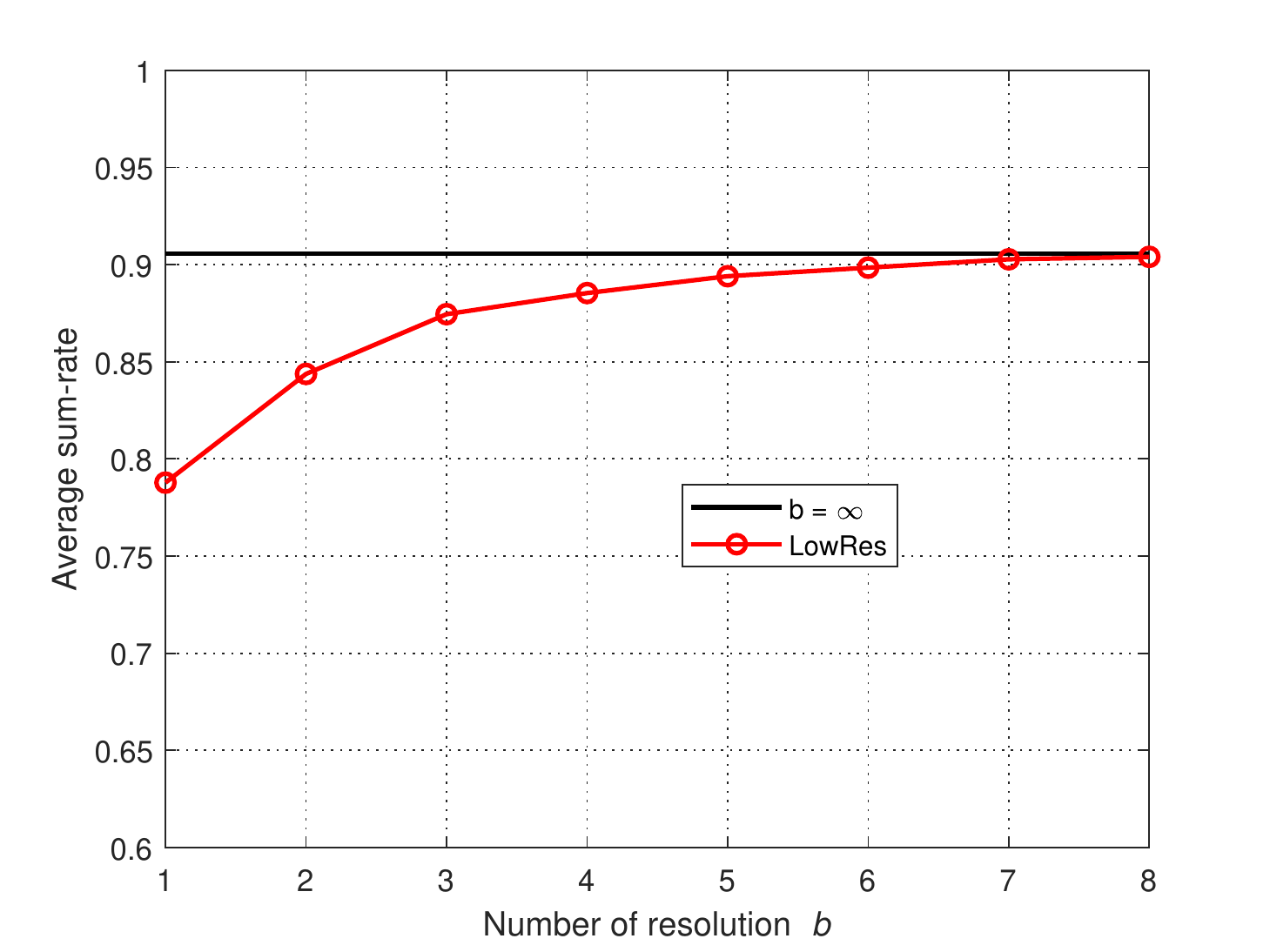}
\caption{Average sum-rate versus the number of iterations ($N_\mathrm{t} = 6$, $M = 64$, $K = 3$, $N = 64$, $N_\mathrm{s} = 4$, $P = -5$ dB).}
\label{fig:asr_vs_b}
\end{figure}

We start with presenting the convergence of the proposed joint beamformer and IRS design by plotting the average sum-rate versus the number of iterations in Fig. \ref{fig:asr_vs_iter}. Simulation results illustrate that the proposed algorithm can converge within 30 iterations when using continuous phase shifters and within 20 iterations when using low-resolution phase shifters to realize the IRS. When the numbers of antennas and IRS elements increase, the proposed algorithm can still converge within limited iterations.
Next in Fig. \ref{fig:asr_vs_b}, we plot the average sum-rate as a function of the resolution $b$ (LowRes) of each IRS element.
Fig. \ref{fig:asr_vs_b} shows that $b=4$ is a sufficiently precise resolution level and the performance improvement is marginal when $b$ is larger than 4.
Moreover, considering the both results of the convergence speed as illustrated in Fig. \ref{fig:asr_vs_iter} and the influence of resolution $b$ as shown in Fig. \ref{fig:asr_vs_b}, it is more practical and efficient to adopt the IRS using low-resolution phase shifters in realistic systems.

\begin{figure}[!t]
\centering
\subfigure[$N_\mathrm{t} = 4$, $M = 64$]{
{\label{fig:asr_vs_p1}}
\includegraphics[height=2.7 in]{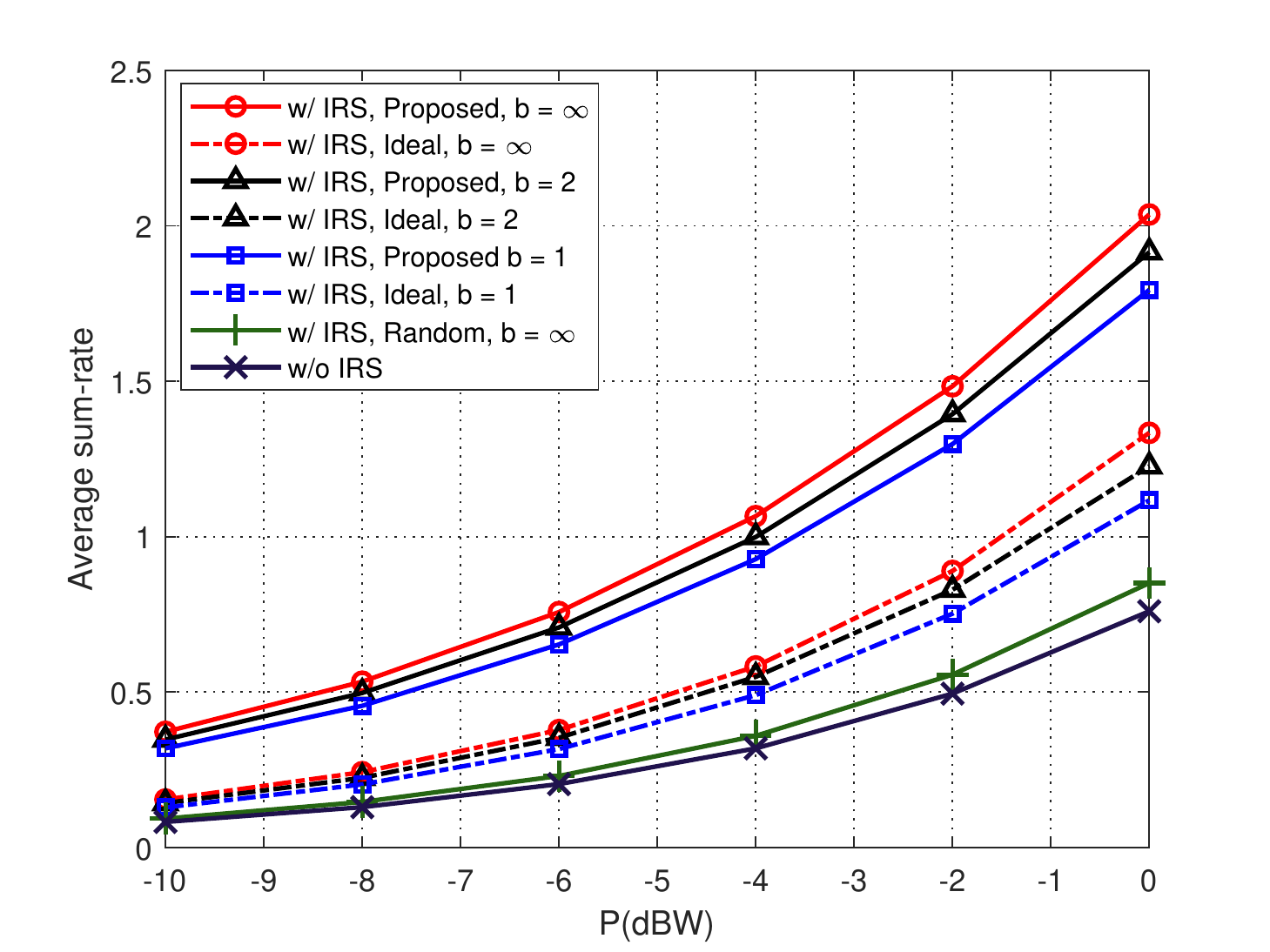}}
\hspace{-0.5cm}
\subfigure[$N_\mathrm{t} = 8$, $M = 144$]{
{\label{fig:asr_vs_p2}}
\includegraphics[height=2.7 in]{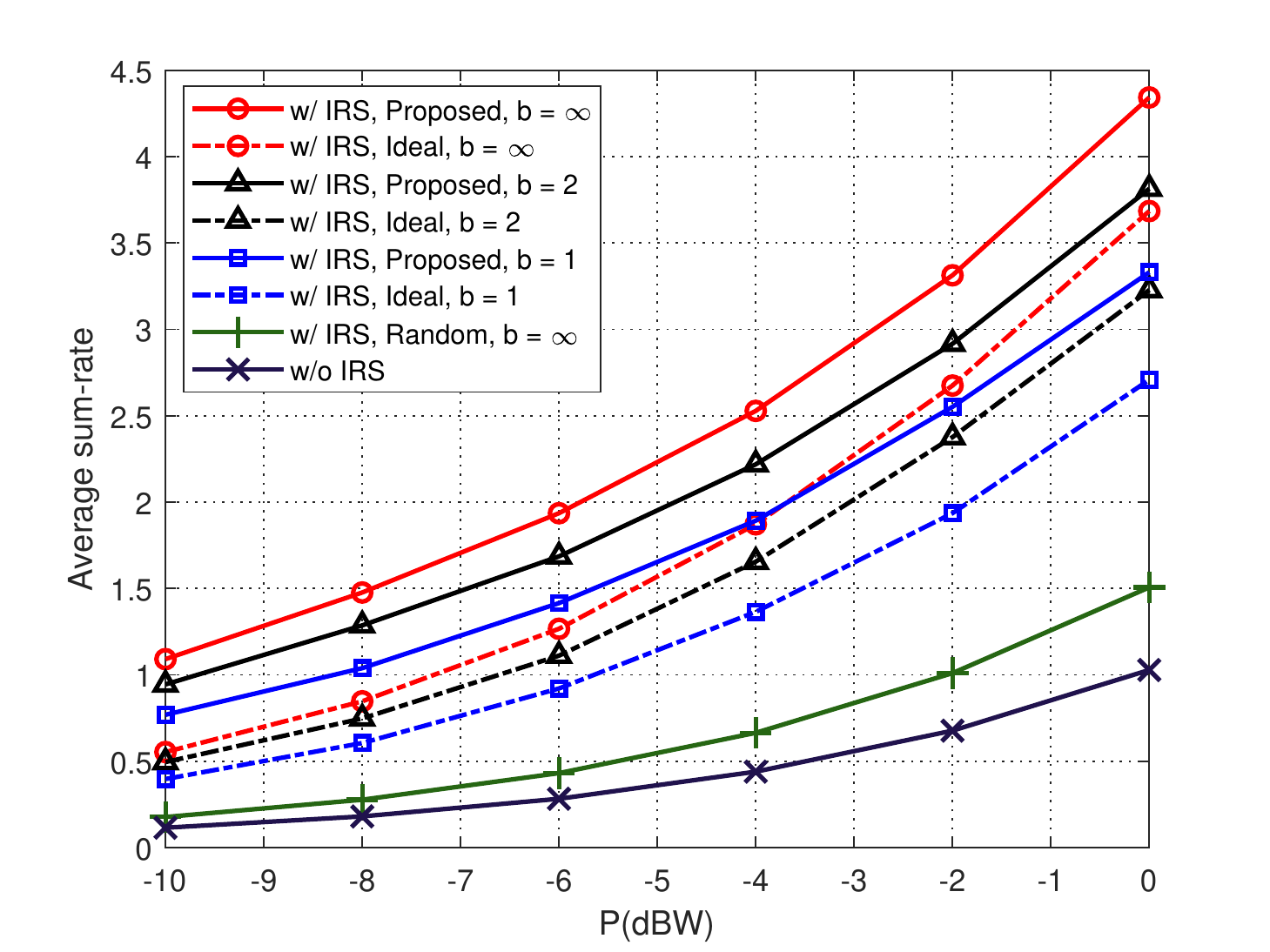}}
\caption{Average sum-rate versus transmit power $P$ ($K = 3$, $N = 64$, $N_\mathrm{s} = 4$).}\label{fig:asr_vs_p}
\end{figure}

Fig. \ref{fig:asr_vs_p} shows the average sum-rate among all subcarriers versus the transmit power $P$ with the proposed algorithm for the cases of using continuous and low-resolution (i.e., $b = 1,2$-bit) phase shifters with different settings (e.g., number of antennas and/or IRS elements).
For fair comparison, we also plot the average sum-rate for the following schemes:
\begin{itemize}
\item The average sum-rate designed by our proposed simplified IRS model in this paper and testified by the same IRS model, which is marked as ``w/ IRS, Proposed''.
\item The average sum-rate designed by the ideal IRS model in \cite{H Li} but testified by the proposed IRS model, which is marked as ``w/ IRS, Ideal''.
\item The average sum-rate designed by the practical two-dimensional IRS model in \cite{S Abeywickrama}, which only considered the impact of reflection amplitude for narrowband systems, but testified by the proposed IRS model, which is marked as ``w/ IRS, Amplitude Only''.
\item Lower bound I: The system with an IRS whose BPSs are randomly selected within the range $[-\pi, \pi]$ and calculated by the proposed IRS model, which is marked as ``w/ IRS, Random''.
\item Lower bound II: The system with direct link only, which is marked as ``w/o, IRS''.
\end{itemize}
It can be observed from Fig. \ref{fig:asr_vs_p} that the proposed algorithm can achieve significantly better performance compared with two lower bounds for all transmit power ranges, which illustrates the advantages of employing IRS in wireless communications.
Moreover, the proposed algorithm also outperforms the ``w/ IRS, Ideal'' scheme, which demonstrates the importance of precisely modeling the reflection characteristics of the practical IRS.
More importantly, from the performance gap between the ``w/ IRS, Proposed'' scheme and the ``w/ IRS, Ideal'' one as well as that between the ``w/ IRS, Proposed'' scheme and the ``w/ IRS, Amplitude Only'' scheme one we can see, although the sum-rate performance is mainly influenced by the amplitude variation of practical IRS, the phase shift variation with respect to different frequencies still, to some extent, causes performance gap, which demonstrates the significance of this work.

\begin{figure}[!t]
\centering
  \includegraphics[width=3.5 in]{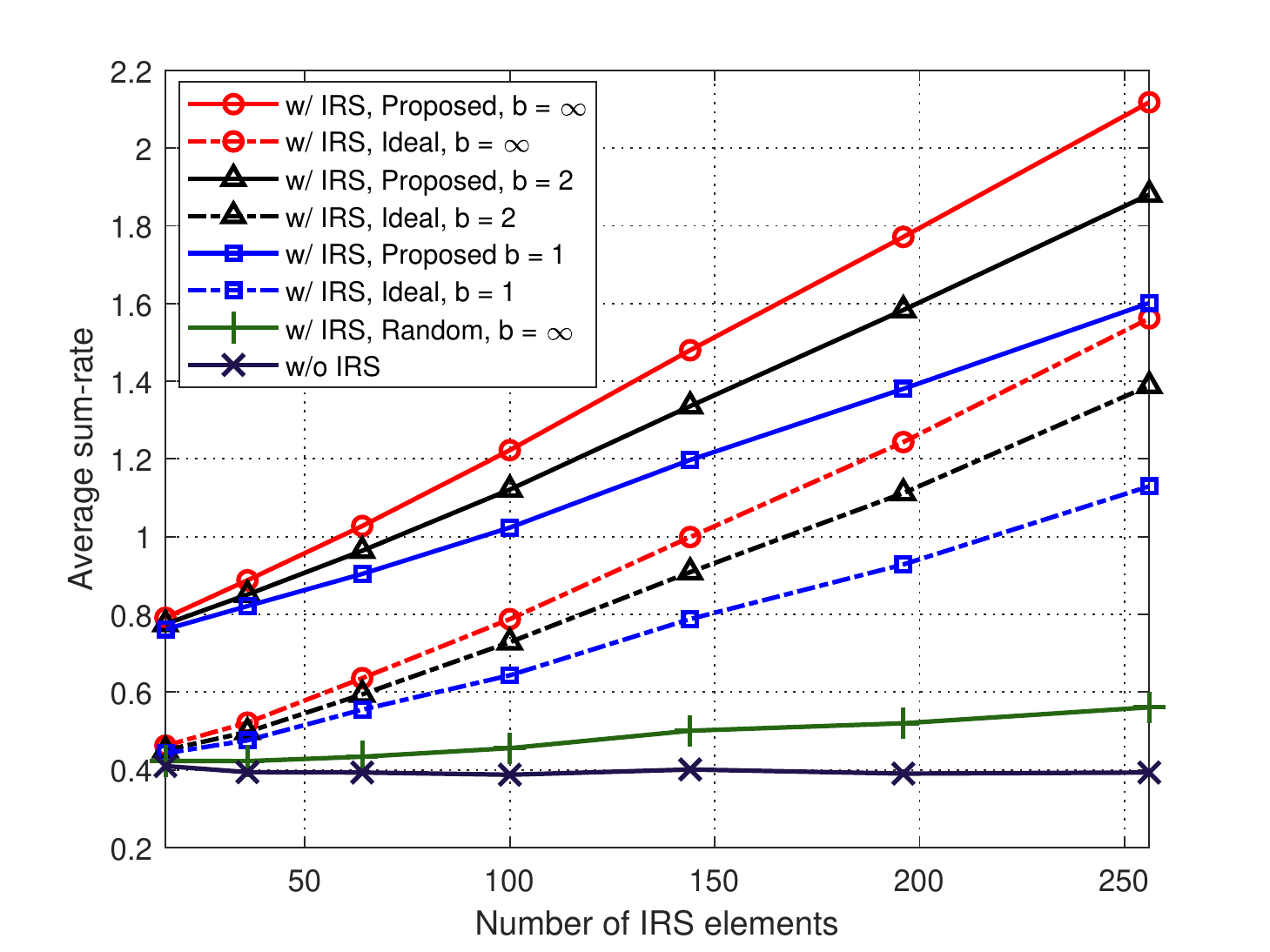}
  \caption{Average sum-rate versus the number of IRS elements $M$ ($N_\textrm{t} = 6$, $K = 3$, $N = 64$, $N_\mathrm{s} = 4$, $P = -5$ dB).}\label{fig:asr_vs_m}
\end{figure}

To illustrate the advantage of employing IRS in enhancing wideband wireless communications, in Fig. \ref{fig:asr_vs_m} we plot the average sum-rate versus different numbers of IRS elements $M$. A similar conclusion can be drawn from Fig. \ref{fig:asr_vs_m} that the proposed algorithm can always achieve better performance than its competitors.
Moreover, with the number of IRS elements growing, the performance gap between the ``w/ IRS'' scheme and the ``w/o IRS'' one is becoming larger.

\begin{figure}[!t]
\centering
  \includegraphics[width=3.5 in]{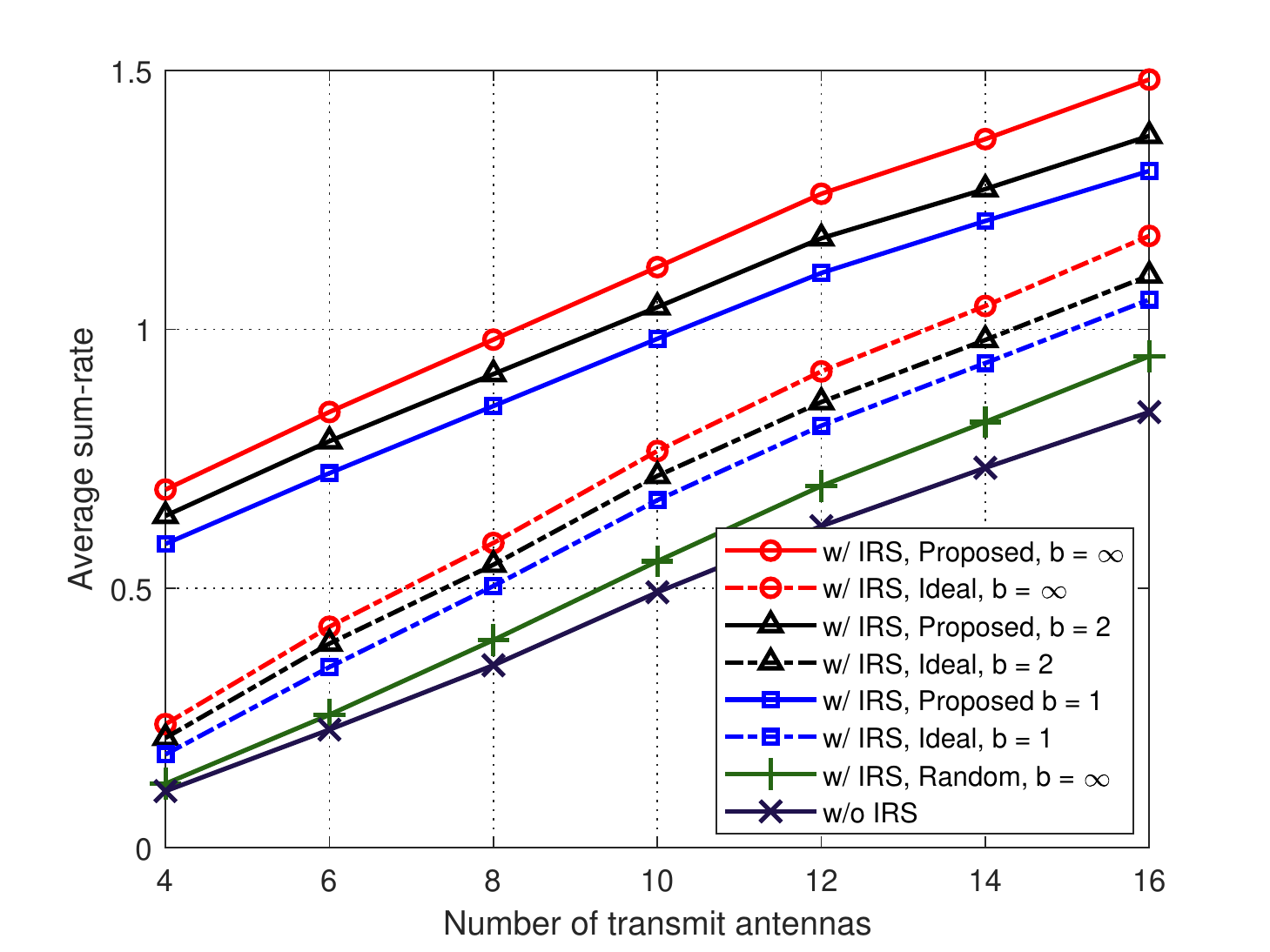}
  \caption{Average sum-rate versus the number of transmit antennas $N_\mathrm{t}$ ($M = 64$, $N = 64$, $N_\mathrm{s} = 8$, $K = 3$, $P = -10$ dB).} \label{fig:asr_vs_nt}
\end{figure}

Finally, the average sum-rate as a function of the number of transmit antennas is illustrated in Fig. \ref{fig:asr_vs_nt}.
A similar conclusion can be obtained from the above simulation results that the proposed ``w/ IRS, Proposed'' scheme always has the best sum-rate performance compared to ``w/ IRS, Amplitude Only/Ideal/Random'' as well as ``w/o IRS'' schemes.

\section{Conclusions}
\label{sc:Conclusions}
In this paper, we first simplified the practical IRS model and validated the accuracy of the proposed model based on numerical simulations.
With the simplified practical model, we considered the problem of joint beamformer and IRS reflection design with both continuous and low-resolution phase shifters to maximize the average sum-rate of a wideband MU-MISO-OFDM system.
We proposed a sub-optimal iterative algorithm by exploiting the equivalence between sum-rate maximization and MSE minimization.
Simulation results demonstrated the significance of modeling the imperfect response characteristics of IRS reflecting elements and its associated configuration design.
With the tremendous difference between the ideal reflection model and the practical reflection model, there are many issues worthy of studying in future works, including IRS deployment, resource allocation, user scheduling, fast channel estimation, as well as learning-based methods, etc.



\begin{thebibliography}{99}






\bibitem{Lee CM 2014} A. L. Swindlehurst, E. Ayanoglu, P. Heydari, and F. Capolino, ``Millimeter-wave massive MIMO: The next wireless revolution?'' \textit{IEEE Commun. Mag.}, vol. 52, no. 9, pp. 56-62, Sept. 2014.

\bibitem{Zhang CST 2017} S. Zhang, Q. Wu, S. Xu, and G. Y. Li, ``Fundamental green tradeoffs: Progress, challenges, and impacts on 5G networks,'' \textit{IEEE Commun. Surveys \& Tutorials}, vol. 19, no. 1, pp. 33-56, First Quarter 2017.

\bibitem{Q Wu 2017} Q. Wu, G. Y. Li, W. Chen, D. W. K. Ng, and R. Schober, ``An overview of sustainable green 5G networks,'' \textit{IEEE Wireless Commun.}, vol. 24, no. 4, pp. 72-80, Aug. 2017.



\bibitem{Q Wu 2019} Q. Wu and R. Zhang, ``Towards smart and reconfigurable environment: Intelligent reflecting surface aided wireless network,'' \textit{IEEE Commun. Mag.}, vol. 58, no. 1, pp. 106-112, Jan. 2020.





\bibitem{Di Renzo 2020} M. Di Renzo, A. Zappone, M. Debbah, M.-S. Alouini, C. Yuen,  J. de Rosny, and S. Tretyakov, ``Smart radio environments empowered by reconfigurable intelligent surfaces: How it works, state of research, and road ahead,'' \textit{IEEE J. Sel. Areas Commun.}, vol. 38, no. 11, pp. 2450-2525, Nov. 2020.



\bibitem{S Gong 2019} S. Gong, \textit{et al}. ``Towards smart radio environment for wireless communications via intelligent reflecting surfaces: A contemporary survey,'' \textit{IEEE Commun. Surveys \& Tutorials}, vol. 22, no. 4, pp. 2283-2314, Fourth Quarter, 2020.


\bibitem{K-K Wong} K-K. Wong, K-F. Tong, Z. Chu, and Y. Zhang, ``A vision to smart radio environment: Surface wave communication superhighways,'' \textit{IEEE Wireless Commun.}, to appear.

\bibitem{Q Wu 2020} Q. Wu, S. Zhang, B. Zheng, C. You, and R. Zhang, ``Intelligent reflecting surface aided wireless communications: A tutorial,'' July 2020. [Online]. Available: https://arxiv.org/abs/2007.02759


\bibitem{C. Huang CM} C. Huang, S. Hu, G. C. Alexandropoulos,  A. Zappone, C. Yuen,  R. Zhang, M. D. Renzo, M. Debbah, ``Holographic MIMO surfaces for 6G wireless networks: Opportunities, challenges, and trends,'' \textit{IEEE Wireless Communications Magazine}, vol. 27, no. 5, pp. 118-125, Oct. 2020.

\bibitem{Y Liu 2020} Y. Liu, X. Liu, X. Mu, T. Hou, J. Xu, Z. Qin, M. D. Renzo, and N. Al-Dhahir, ``Reconfigurable intelligent surfaces: Principles and opportunities,'' July 2020. [Online]. Available: https://arxiv.org/abs/2007.03435

\bibitem{N Rajatheva} N. Rajatheva, \textit{et al.}, ``White paper on broadband connectivity in 6G,'' Apr. 2020. [Online]. Available: https://arxiv.org/abs/2004.14247



\bibitem{X Yu 2019} X. Yu, D. Xu, and R. Scholar, ``MISO wireless communication systems via intelligent reflecting surface,'' in \textit{Proc. IEEE Int. Conf. Commun. China (ICCC)}, Changchun, China, Dec. 2019.

\bibitem{R Zhang} Q. Wu and R. Zhang, ``Intelligent reflecting surface enhanced wireless network via joint active and passive beamforming,'' \textit{IEEE Trans. Wireless Commun.}, vol. 18, no. 11, pp. 5394-5409, Nov. 2019.

\bibitem{Y Han} Y. Han, W. Tang, S. Jin, C. Wen, and X. Ma, ``Large intelligent surface-assisted wireless communication exploiting statistical CSI,'' \textit{IEEE Trans. Veh. Technol.}, vol. 68, no. 8, pp. 8238-8242, Aug. 2019.

\bibitem{C Huang} C. Huang, A. Zappone, G. C. Alexandropoulos, M. Debbah, and C. Yuen, ``Reconfigurable intelligent surfaces for energy efficiency in wireless communication,'' \textit{IEEE Trans. Wireless Commun.}, vol. 18, no. 8, pp. 4157-4170, Aug. 2019.


\bibitem{C Huang JSAC} C. Huang, R. Mo, and C. Yuen, ``Reconfigurable intelligent surface assisted multiuser MISO systems exploiting deep reinforcement learning,'' \textit{IEEE J. Sel. Area Commun. (JSAC)}, vol. 38, no. 8, pp. 1839-1850.


\bibitem{H Guo} H. Guo, Y.-C. Liang, J. Chen, and E. G. Larsson, ``Weighted sum-rate optimization for intelligent reflecting surface enhanced wireless networks,'' in \textit{Proc. IEEE Global Commun. Conf. (GLOBECOM)}, Waikoloa, HI, Dec. 2019.

\bibitem{J Zhao} J. Zhao, ``Optimizations with intelligent reflecting surfaces (IRSs) in 6G wireless networks: Power control, quality of service, max-min fair beamforming for unicast, broadcast, and multicast with multi-antenna mobile users and multiple IRSs,'' Aug. 2019. [Online]. Avaliable: https://arXiv.org/abs/1908.03965

\bibitem{M Li 2020} Y. Liu, J. Zhao, M. Li, and Q. Wu, ``Intelligent reflecting surface aided MISO uplink communication network: Feasibility and power minimization for perfect and imperfect CSI,'' July 2020. [Online]. Avaliable: https://arXiv.org/abs/2007.01482

\bibitem{B Di} B. Di, H. Zhang, L. Song, Y. Li, Z. Han, and H. V. Poor, ``Hybrid beamforming for reconfigurable intelligent surface based multi-user communications: Achievable rates with limited discrete phase shifts,'' \textit{IEEE J. Sel. Area Commun.}, vol. 38, no. 8, pp. 1809-1822, Aug. 2020.

\bibitem{Q Wu} Q. Wu and R. Zhang, ``Beamforming optimization for wireless network aided by intelligent reflecting surface with discrete phase shifts,'' \textit{IEEE Trans. Commun.}, vol. 68, no. 3, pp. 1838-1851, Mar. 2020.

\bibitem{J Xu} J. Xu, W. Xu, and A. L. Swindlehurst, ``Discrete phase shift design for practical large intelligent surface communication,'' in \textit{Proc. IEEE Pacific Rim Conf. on Commun., Computers and Signal Process. (PACRIM)}, Victoria, Canada, Aug. 2019.

\bibitem{J He} J. He, K. Yu, and Y. Shi, ``Coordinated passive beamforming for distributed intelligent reflecting surfaces network'', in \textit{Proc. IEEE Veh. Technol. Conf. (VTC)}, Virtual Conference, May 2020.

\bibitem{Z Li} Z. Li, M. Hua, Q. Wang, and Q. Song, ``Weighted sum-rate maximization for multi-IRS aided cooperative transmission'', \textit{IEEE Wireless Commun. Lett.}, vol. 9, no. 10, pp. 1620-1624, Oct. 2020.

\bibitem{B Zheng Multi-IRS} B. Zheng, C. You, and R. Zhang, ``Double-IRS assisted multi-user MIMO: Cooperative passive beamforming design'', Nov. 2020. [Online]. Avaliable: https://arxiv.org/abs/2008.13701

\bibitem{M Cui} M. Cui, G. Zhang, and R. Zhang, ``Secure wireless communication via intelligent reflecting surface,'' \textit{IEEE Wireless Commun. Lett.}, vol. 8, no. 5, pp. 1410-1414, Oct. 2019.

\bibitem{X Guan Jan} X. Guan, Q. Wu, and R. Zhang, ``Intelligent reflecting surface assisted secrecy communication: Is artificial noise helpful or not?'' \textit{IEEE Wireless Commun. Lett.}, vol. 9, no. 6, pp. 778-782, June 2020.

\bibitem{D Xu} D. Xu, X. Yu, Y. Sun, D. W. K. Ng, and R. Schober, ``Resource allocation for secure IRS-assisted multiuser MISO systems,'' in \textit{Proc. IEEE Global Commun. Conf. (GLOBECOM)}, Waikoloa, HI, Dec. 2019.

\bibitem{L Zhang} L. Zhang, C. Pan, Y. Wang, H. Ren, K. Wang, and A. Nallanathan, ``Robust beamforming design for intelligent reflecting surface aided cognitive radio systems with imperfect cascaded CSI,'' Apr. 2020. [Online]. Avaliable: https://arxiv.org/abs/2004.04095

\bibitem{D Xu 2020} D. Xu, X. Yu, and R. Schober, ``Resource allocation for intelligent reflecting surface-assisted cognitive radio networks,'' in \textit{IEEE Signal Process. Advances Wireless Commun. (SPAWC)}, Virtual Conference, May 2020.

\bibitem{X Guan} X. Guan, Q. Wu, and R. Zhang, ``Joint power control and passive beamforming in IRS-assisted spectrum sharing,'' \textit{IEEE Commun. Lett.}, vol. 24, no. 7, pp. 1553-1557, July 2020.

\bibitem{A Khaleel} A. Khaleel and E. Basar, ``Reconfigurable intelligent surface-empowered MIMO systems,'' \textit{IEEE Syst. J.}, to appear.

\bibitem{E Basar} E. Basar, ``Reconfigurable intelligent surface-based index modulation: A new beyond MIMO paradigim for 6G,'' \textit{IEEE Trans. Commun.}, vol. 68, no. 5, pp. 3187-3196, May 2020.

\bibitem{B Zheng NOMA} B. Zheng, Q. Wu, and R. Zhang, ``Intelligent reflecting surface-assisted multiple access with user pairing: NOMA or OMA?'' \textit{IEEE Commun. Lett.}, vol. 24, no. 4, pp. 753-757, Apr. 2020.

\bibitem{Y Yang} Y. Yang, B. Zheng, S. Zhang, and R. Zhang, ``Intelligent reflecting surface meets OFDM: Protocal design and rate maximization,'' \textit{IEEE Trans. Commun.}, vol. 68, no. 7, pp. 4522-4535, July 2020.

\bibitem{B Zheng} B. Zheng and R. Zhang, ``Intelligent reflecting surface-enhanced OFDM: Channel estimation and reflection optimization,'' \textit{IEEE Wireless Commun. Lett.}, vol. 9, no. 4, pp. 518-522, Apr. 2020.

\bibitem{T Bai} T. Bai, C. Pan, H. Ren, Y. Deng, M. Elkashlan, and A. Nallanathan, ``Resource allocation for intelligent reflecting surface aided wireless powered mobile edge computing in OFDM systems,'' Mar. 2020. [Online]. Available: https://arxiv.org/abs/2003.05511

\bibitem{H Li} H. Li, R. Liu, M. Li, Q. Liu, and X. Li, ``IRS-enhanced widebadn MU-MISO-OFDM communication systems,'' in \textit{Proc. IEEE Wireless Commun. Networking Conf. (WCNC)}, Seoul, South Korea, May 2020.


\bibitem{H Rajagopalan} H. Rajagopalan and Y. Rahmat-Samii, ``Loss quantification for microstrip reflectarray: Issue of high fields and currents,'' in \textit{Proc. IEEE Antennas and Propag. Society Int. Symposium}, San Diego, CA, July 2008.

\bibitem{W Tang} W. Tang \textit{et al.}, ``MIMO transmission through reconfigurable intelligent surface: System design, analysis, and implementation,'' \textit{IEEE J. Sel. Areas Commun.}, vol. 38, no. 11, pp. 2683-2699, Nov. 2020.


\bibitem{S Abeywickrama} S. Abeywickrama, R. Zhang, Q. Wu, and C. Yuen, ``Intelligent reflecting surface: Practical phase shift model and beamforming optimization,'' \textit{IEEE Trans. Commun.}, vol. 68, no. 9, pp. 5849-5863, Sept. 2020.

\bibitem{W Cai} W. Cai, H. Li, M. Li, and Q. Liu, ``Practical modeling and beamforming for intelligent reflecting surface aided wideband systems,'' \textit{IEEE Commun. Lett.}, vol. 24, no. 7, pp. 1568-1571, July 2020.

\bibitem{S Koziel} S. Koziel and L. Leifsson, \textit{Surrogate-based modeling and optimization}. Springer, 2013.



\bibitem{Taha} A. Taha, M. Alrabeiah, and A. Alkhateeb, ``Enabling large intelligent surfaces with compressive sensing and deep learning,'' Apr. 2019. [Online]. Avaliable: https://arxiv.org/abs/1904.10136


\bibitem{B Zheng 2020} B. Zheng, C. You, and R. Zhang, ``Intelligent reflecting surface assisted multi-user OFDMA: Channel estimation and training design,'' \textit{IEEE Trans. Wireless Commun.}, to appear.

\bibitem{B Zheng OFDM CE} B. Zheng, C. You, and R. Zhang, ``Fast channel estimation for IRS-assisted OFDM,'' \textit{IEEE Wireless Commun. Lett.}, to appear.

\bibitem{W. Yang} W. Yang, H. Li, M. Li, Y. Liu, and Q. Liu, ``Channel estimation for practical IRS-assisted OFDM systems,'' Dec. 2020. [Online]. Available: https://arxiv.org/abs/2012.13521


\bibitem{SPAWC 2017} Y. Kwon, J. Chung, and Y. Sung, ``Hybrid beamformer design for mmWave wideband multi-user MIMO-OFDM systems,'' in \textit{Proc. IEEE Int. Workshop on Signal Process. Advances in Wireless Commun. (SPAWC)}, Sapporo, Japan, July 2017.


\bibitem{Q Shi 2011} Q. Shi, M. Razaviyayn, Z. Q. Luo, and C. He, ``An iteratively weighted MMSE approach to distributed sum-utility maximization for a MIMO interfering broadcast channel,'' \textit{IEEE Trans. Signal Process.}, vol. 59, no. 9, pp. 4331-4340, Sept. 2011.

\bibitem{Bertsekas 1999} D. Bertsekas, \textit{Nonlinear Programming}, 2nd ed. Belmont, MA, USA: Athena Scientific, 1999.

\end{thebibliography}
\end{document}